\documentclass[bibyear]{aa}
\usepackage[varg]{txfonts}
\usepackage{graphicx}
\def\G{\mbox{G31.41+0.31}}

\def\Msun{\mbox{$M_\odot$}}
\def\Lsun{\mbox{$L_\odot$}}

\def\HCOp{HCO$^+$}
\def\HCOpI{H$^{13}$CO$^+$}
\def\CO{$^{12}$CO}

\def\CSI{\mbox{C$^{34}$S}}
\def\CSII{\mbox{C$^{33}$S}}

\def\MCN{\mbox{CH$_3$CN}}

\def\MAC{\mbox{CH$_3$CCH}}

\def\HM{\mbox{H$_2$}}
\def\HII{H{\sc ii}}

\def\UC{UC~H{\sc ii}}

\def\kms{\mbox{km~s$^{-1}$}}
\def\cmc{cm$^{-3}$}
\def\cmq{cm$^{-2}$}

\def\mic{\mbox{$\mu$m}}

\def\Te{\mbox{$T_{\rm e}$}}
\def\ne{\mbox{$n_{\rm e}$}}
\def\Vt{V_{\rm t}}
\def\Tt{T_{\rm t}}
\def\Trot{\mbox{$T_{\rm rot}$}}

\def\Ncol{\mbox{$N_{\rm CH_3CCH}$}}

\begin{document}
\title{
The GUAPOS project
}
\subtitle{
VII: Physical structure and molecular environment of the \G\ \HII\ region\thanks{Based on observations carried out with ALMA and the VLA}
}
\author{
        R.~Cesaroni\inst{1} \and
	M. T. Beltr\'an\inst{1} \and
	V.~M.~Rivilla\inst{2} \and
	\'A.~S\'anchez-Monge\inst{3,4} \and
	L.~Colzi\inst{2} \and F.~Fontani\inst{1} \and
	\'A. L\'opez-Gallifa\inst{2} \and \\
	A. Lorenzani\inst{1} \and
	C. Mininni\inst{5}
}
\institute{
 INAF, Osservatorio Astrofisico di Arcetri, Largo E. Fermi 5, I-50125 Firenze, Italy
	   \email{riccardo.cesaroni@inaf.it}
 \and
 Centro de Astrobiolog\'{\i}a (CAB), CSIC-INTA, Carretera de Ajalvir km 4, Torrej\'on de Ardoz, E-28850 Madrid, Spain
 \and
 Institut de Ci\`encies de l'Espai (ICE), CSIC, Campus UAB, Carrer de Can Magrans s/n, E-08193, Bellaterra, Barcelona, Spain
 \and
 Institut d'Estudis Espacials de Catalunya (IEEC), E-08860, Castelldefels, Barcelona, Spain
 \and
 INAF, Istituto di Astrofisica e Planetologia Spaziale, Via Fosso del
Cavaliere 100, I-00133 Roma, Italy
}
\offprints{R. Cesaroni, \email{riccardo.cesaroni@inaf.it}}
\date{Received date / Accepted date}

\abstract{
Ionised regions around OB-type stars are formed at an early stage of their
evolution and are important to investigate the formation process of these
objects. However, so far only few observations of their physical structure
and interaction with the parental molecular cloud have been made. The
high resolution and sensitivity of new instruments such as ALMA and the
upgraded VLA allow us to fill this gap in our knowledge.
}{
We investigate the well known core-halo 
ultracompact
\HII\ region \G\ and the surrounding
molecular clump with the aim to determine the density and temperature
of both the ionised and neutral gas, and possibly obtain a 3D picture
of their spacial distribution.
}{
We take advantage of the full-band frequency coverage at 3~mm obtained
with
ALMA for
the GUAPOS project to image the emission of a plethora of hydrogen
recombination lines towards the \G\ \HII\ region as well as several
molecular transitions which are tracers of medium-density
($\sim$$10^4$--$10^6$~\cmc)
gas.
The line data are complemented by continuum measurements obtained with
the VLA at 1~cm and 7~mm.
By fitting
these lines also using a model that takes into account non-LTE effects we
can
investigate the density and temperature structure and
the velocity field of the region.
}{
Our findings,
based on a model fit accounting for non-LTE effects,
indicate that the electron temperature of the \HII\ region is
mostly spanning a range between 5000 and 6000~K, while the density varies
between 2500 and 7500~\cmc. All in all, the distribution of these parameters
as well as the corresponding velocity field hint at a cometary shaped
\HII\ region expanding away from the observer to the NW. The molecular
gas appears to be still infalling towards the peak of the \UC\ region,
and its density and temperature are consistent with pressure confinement
of the ionised gas to the SE.
}
{}

\maketitle

\section{Introduction}
\label{sint}

Ultracompact (UC) \HII\ regions are early manifestations of newly formed
early-type stars, which ionise the surrounding environment with their
powerful Lyman continuum flux, shortward of 912~\AA. The pivotal Very
Large Array (VLA) survey by Wood \& Churchwell (\cite{wc89a}) and their
subsequent study on the far-IR colours of \UC\ regions (Wood \& Churchwell
\cite{wc89b}) made it possible to identify a conspicuous number of these
sources and propose a classification based on their morphology. Since
then, a number of observations have been performed to investigate the
structure and evolution of these intriguing objects, but the limited
sensitivity, {\it uv} coverage, and frequency coverage of the available
interferometers -- especially at (sub)millimeter wavelengths -- hindered
the simultaneous observation of both the ionised gas and its molecular
surroundings.
Measuring at the same time many different tracers allows to obtain very
accurate relative positions and intensities, which are instead affected
by significant uncertainties when comparing data obtained with different
instruments and/or at different times.
Investigating not only the structure of the ionised and molecular
components, but also their mutual interaction is important to understand
the evolution of \UC\ regions in such a complex environment, as suggested
by theoretical studies (e.g. Peters et al.~\cite{pete10a,pete10b,pete10c}).

The advent of new-generation instruments such as the upgraded Karl Jansky
Very Large Array (VLA) and the Atacama Large Millimeter and submillimeter
Array (ALMA) has dramatically improved the situation. It is now possible
to cover broad frequency ranges in a reasonable observing time, thus
obtaining simultaneous measurements in many different molecular tracers and
recombination lines
(tracing the ionised gas)
as well as high-sensitivity continuum images. With this
in mind, we decided to study in depth the well known high-mass star forming
region \G, located at a distance of 3.75~kpc (Immer et al. \cite{immer19})
and containing an \UC\ region, classified as ``core-halo'' by
Wood \& Churchwell (\cite{wc89a}), and a bright hot molecular core
(HMC). The latter has been the subject of a series of articles aiming at
investigating its physical structure and chemical composition (Cesaroni
et al.~\cite{cesa94a,cesa94b,cesa98,cesa10,cesa11,cesa17}; Beltr\'an et
al.~\cite{belt04,belt05,belt09,belt18,belt19,belt21,belt22a,belt22b,belt24};
Rivilla et al.~\cite{rivi17}).  The goal of our ALMA project ``G31.41+0.31
Unbiased ALMA sPectral Observational Survey'' (GUAPOS) was to cover the whole
bandwidth of the 3~mm receivers of the interferometer, which allowed the
detection of a plethora of transitions from a variety of molecular species
(Mininni et al.~\cite{mini20,mini23}; Colzi et al.~\cite{colz21}; Fontani
et al.~\cite{font24}; L\'opez-Gallifa et al.~\cite{loga24,loga25}). While
this was intended to be mostly a line survey to study the rich chemistry
of HMCs, the detection of many hydrogen recombination lines and several
rotational transitions tracing the extended molecular gas allows us to
shed light also on the internal physical structure of the \HII\ region
and its parental molecular clump.

In this article we present the observations performed with ALMA and the
VLA (Sect.~\ref{sobs}), describe the results obtained (Sect.~\ref{sres}),
derive a number of physical parameters of the ionised and neutral gas
(Sect.~\ref{sana}), and eventually examine the complex structure of the
region with a special focus on the interaction between the \HII\ region and
its molecular surroundings (Sect.~\ref{sdis}).

\section{Observations and data reduction}
\label{sobs}

In the following we describe the observations performed towards \G\ with the ALMA
and VLA interferometers, including also data from the VLA archive.

\subsection{Atacama Large Millimeter and submillimeter Array}
\label{salma}

The data used in this study are part of the GUAPOS project
(2017.1.00501.S, P.I. M.~T. Beltr\'an) and the reader may refer to Mininni
et al. (\cite{mini20}) for the observational details and the data reduction
procedures adopted. Here, we only provide the basic information.

The ALMA observations of \G\ were performed in band~3 with a
phase center of $\alpha$(ICRS)=$18^{\rm h}47^{\rm m}34\fs315$ and
$\delta$(ICRS)=\mbox{--01\degr}12\arcmin45\farcs9. Source J1751+0939 was
used as bandpass and flux calibrator, while the phase calibrator
was J1851+0035. Nine correlator configurations, each consisting of
four units of 1.875~GHz, were necessary to cover the whole receiver
bandwidth between 84.05 and 115.91~GHz, with a spectral resolution
of 0.488~MHz, corresponding to $\sim$1.3--1.7~\kms\ depending on the
frequency. Calibration and data reduction were performed with the
Common Astronomy Software Applications\footnote{https://casa.nrao.edu}
(CASA; McMullin et al.~\cite{mcmu07}) and the maps were all
cleaned with Briggs {\em robust=0.5} weighting and a circular
synthesised beam with full width at half power (FWHP) of 1\farcs2,
corresponding to 4500~au. The continuum maps and corresponding
continuum-subtracted channel maps were created with the software
STATCONT\footnote{https://hera.ph1.uni-koeln.de/$\sim$sanchez/statcont.html}
(S\'anchez-Monge et al.~\cite{statcont}).

\subsection{Karl Jansky Very Large Array}

The ${\rm K_a}$ band observations of \G\ (project 16A-181,
P.I. V.~M.~Rivilla) were performed on March 3, 8, and 22, 2016 in
the C-array configuration. The equatorial coordinates of the
phase center were $\alpha$(J2000)=$18^{\rm h}47^{\rm m}34\fs315$ and
$\delta$(J2000)=\mbox{--01\degr}12\arcmin45\farcs9. The total observing bandwidth
(per polarization) used to measure the continuum emission was $\sim$1.8~GHz.

The primary flux calibrator and the phase calibrator were, respectively,
3C286 and J1851+0035. The CASA software was used for calibration and data
reduction and the data were calibrated with the VLA pipeline. The maps
were made with task {\em tclean} of CASA using Briggs {\em robust=0.5}
weighting. For the sake of comparison with the ALMA data, we also
selected the {\it uv} range 4.9--257~klambda and a circular clean beam
with FWHP=1\farcs2. The resulting RMS noise was $\sim$0.1~mJy/beam.

\begin{figure}
\centering
\resizebox{8.5cm}{!}{\includegraphics[angle=0]{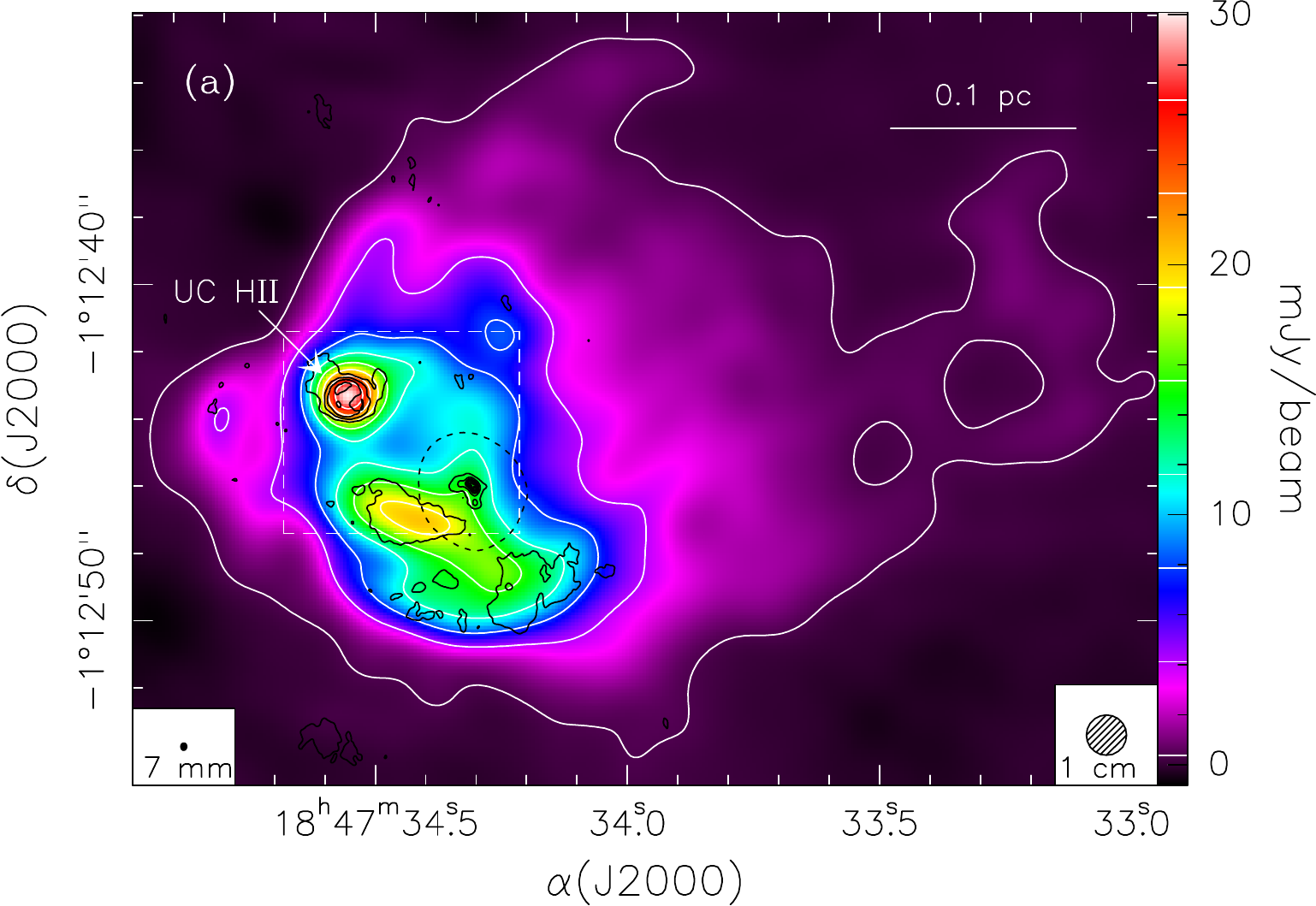}} \\
\resizebox{8.5cm}{!}{\includegraphics[angle=0]{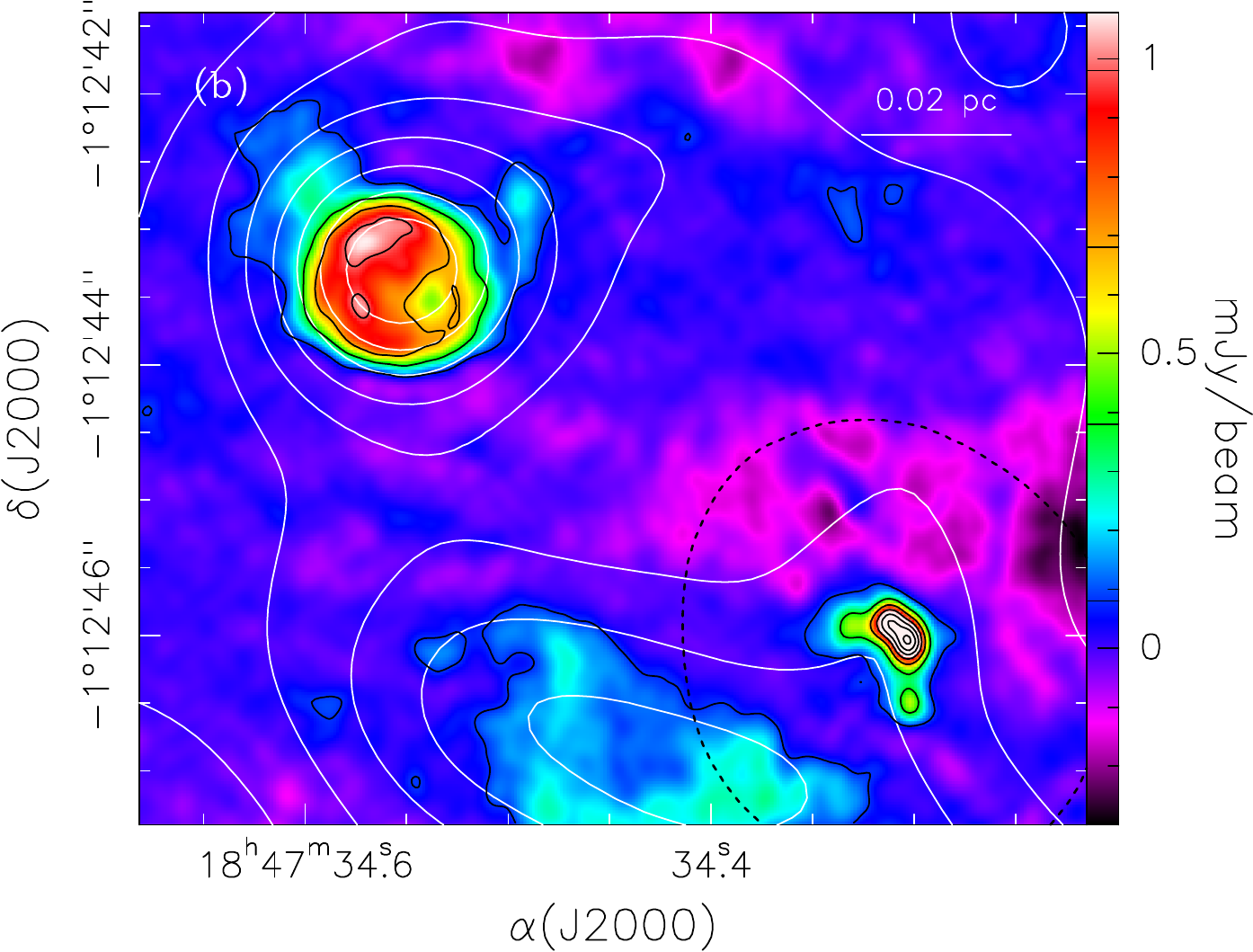}} \\
\resizebox{8.5cm}{!}{\includegraphics[angle=0]{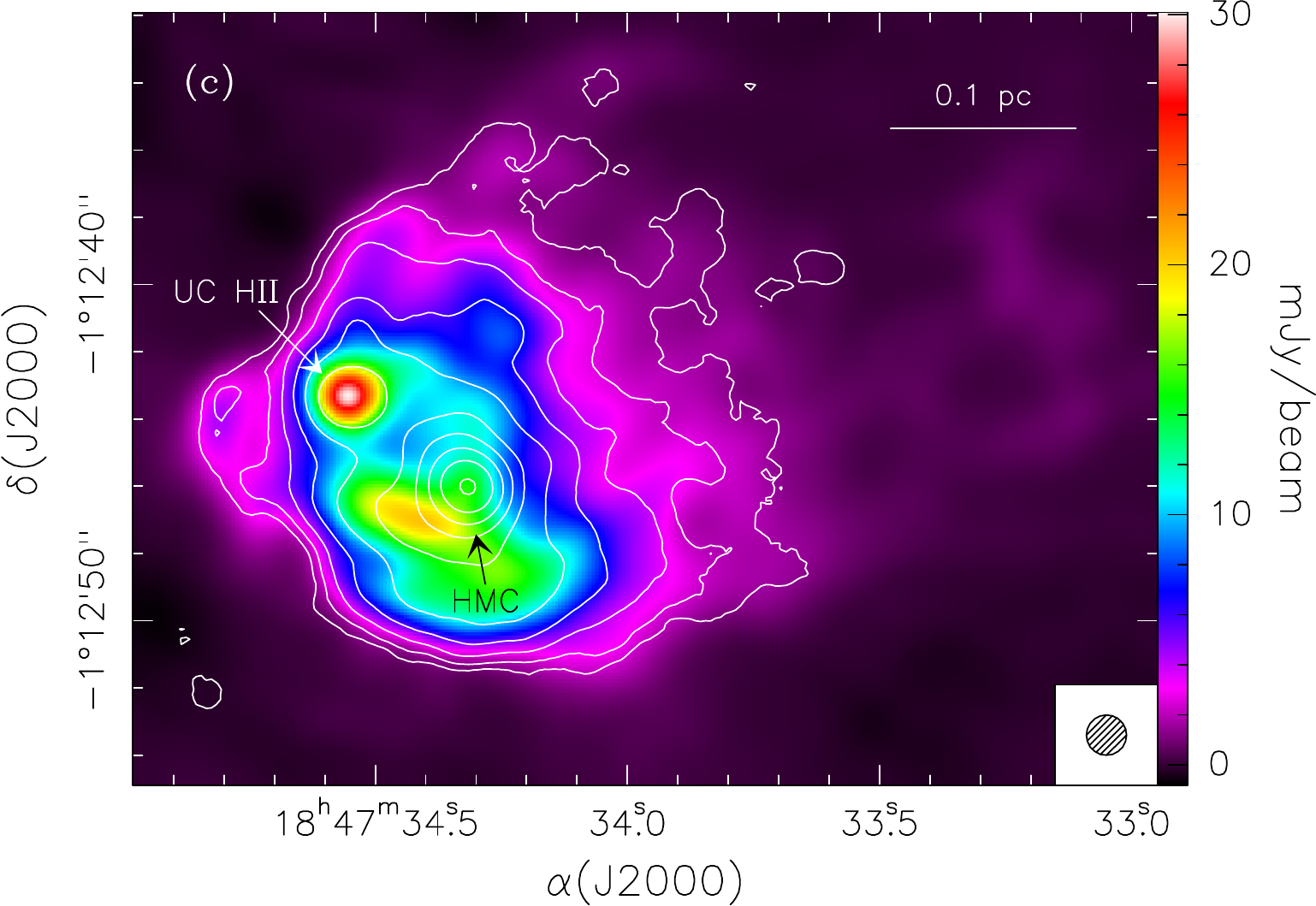}}
\caption{
{\bf a.} Maps of the 1~cm (colour image and white contours) and 7~mm (black
contours) continuum emission imaged with the VLA. The contour levels of
the 1~cm map are drawn in the colour scale to the right, while those of
the 7~mm map range from 0.08 to 1.88 in steps of 0.3~mJy/beam. The black
dotted pattern outlines the approximate border of the HMC (see text). The
dashed rectangle frames the region shown in panel~b. The synthesised beams
(1\farcs2 at 1~cm and 0\farcs21$\times$0\farcs17 with PA --0\fdg5 at 7~mm)
are shown in the bottom left and right corners.
{\bf b.} Enlargement of the region that contains the \UC\ region and the
HMC, corresponding to the dashed rectangle in panel~a. The symbols have
the same meaning as in panel a, with the exception of the colour image
that shows the 7~mm continuum emission. We note that the colour scale is
saturated (the peak emission at 7~mm is 1.96~mJy/beam) to emphasise the
structure of the \UC\ region.
{\bf c.} Contour map of the 3~mm continuum emission obtained from the ALMA data
overlaid on the same 1~cm continuum image as in panel a.
Contour levels range from 0.5 to 180~mJy/beam in 10 logarithmic steps.
The synthesised beam (1\farcs2) is the same for both maps and is shown
in the bottom right corner.
}
\label{fvmaps}
\end{figure}

We also made use of archival data at 7~mm (Project~23A-066,
P.I.~H.~Liu). In this case the observations were performed in the
B-array configuration on January 14 and May 3, 6, 9, 10, 12 2023. The
phase center was $\alpha$(J2000)=$18^{\rm h}47^{\rm m}34\fs308$ and
$\delta$(J2000)=\mbox{--01\degr}12\arcmin45\farcs9. The total observing bandwidth
(per polarization) used to measure the continuum emission was $\sim$7~GHz.
The flux and phase calibrator swere, respectively, 3C286 and J1851+0035. The calibrated data were obtained
directly from the NRAO archive and the continuum maps were made with the CASA
software using task {\em tclean} with Briggs {\em robust=0.5} weighting. The
resulting image has a synthesised beam FWHP of 0\farcs210$\times$0\farcs169
and PA=--0\fdg51. The 1$\sigma$ RMS noise is $\sim$0.01~mJy/beam.

\section{Results}
\label{sres}

In the following, we present the outcome of our observations separately
for the continuum, recombination lines, and molecular lines.

\subsection{Continuum emission}
\label{shii}

The 1~cm and 7~mm continuum maps shown in Fig.~\ref{fvmaps}a trace the
free-free emission from the ionised gas, confirming the core-halo morphology
identified by Wood \& Churchwell~(\cite{wc89a}). Hereafter, we will refer
to the ``core'' as the ``\UC\ region'' to distinguish it from the ``halo'',
i.e. the extended ionised component, and we will use the more generic term
``\HII\ region'' to indicate the whole core+halo system. In particular,
the sub-arcsecond resolution of the 7~mm image reveals the inner structure
of the \UC\ region, which presents an asymmetry hinting at a cometary shape
(see Fig.~\ref{fvmaps}b) with the head pointing approximately eastwards.
Another interesting feature seen at 7~mm is the presence of four emission
peaks inside the HMC, whose approximate border is outlined by the dotted
contour in the figure. The latter corresponds to the 10\% contour level of the
\CSII(2--1) integrated emission (see Sects.~\ref{smolin}
and~\ref{smolc}). These sources correspond to those revealed by Cesaroni
et al.~(\cite{cesa10}) and Beltr\'an et al.~(\cite{belt21}) and may be
associated with newly formed deeply embedded massive stars, as discussed
by these authors.

From the ALMA data we obtained a continuum map by averaging the
emission over the frequency range covered by the GUAPOS project after
removing the line emission as explained in Sect.~\ref{salma}. In
the 3~mm map presented in Fig.~\ref{fvmaps}c one clearly sees a pronounced peak
of emission corresponding to the HMC, which proves that the observed
flux is mostly contributed by dust thermal emission.

\begin{figure}
\centering
\resizebox{8.5cm}{!}{\includegraphics[angle=0]{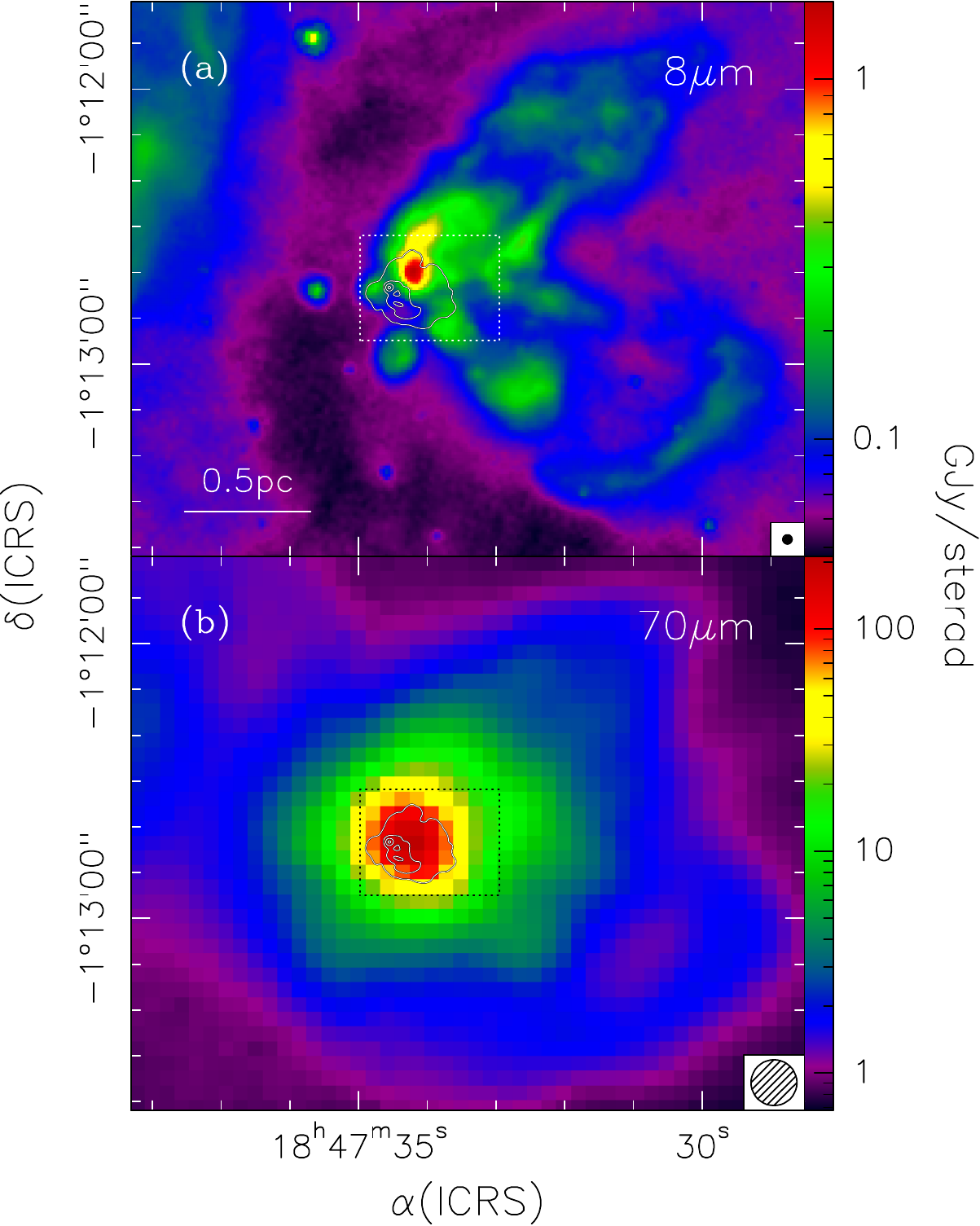}}
\caption{
{\bf a.} Overlay of the 1~cm map of Fig.~\ref{fvmaps} (contours) and the
8~\mic\ image from the Spitzer/GLIMPSE database (colour image). Contour
levels range from 1 to 28 in steps of 9~mJy/beam. The angular resolution of
the latter image is shown in the bottom right corner. The dotted rectangle
outlines the region shown in Figs.~\ref{fvmaps}a and~\ref{fvmaps}c.
{\bf b.} Same as panel a, for the 70~\mic\ image obtained from the
Herschel/Hi-GAL survey.
}
\label{fhigal}
\end{figure}

It is interesting to compare our maps with far-IR images tracing the
emission from the dusty molecular clump enshrouding the \HII\ region.
Figure~\ref{fhigal} presents an overlay between the 1~cm map and
images of the region at 8~\mic\ from the {\it Spitzer}/GLIMPSE database
(Werner et al.~\cite{werner}; Fazio et al.~\cite{fazio}; Benjamin et
al.~\cite{glimpse}) and at 70~\mic\ from the {\it Herschel}/Hi-GAL survey
(Molinari et al.~\cite{higal}). The complexity of the region shows up
clearly in this figure, where the ionised gas appears to be tightly confined
due to the presence of a molecular cloud on the eastern side revealed by the
heavily extincted region elongated in the N--S direction and traced also by
the N$_2$H$^+$(1--0) line emission (see Beltr\'an et al.~\cite{belt22b}).
Noticeably, the HMC is undetected at least up to 8~\mic, despite its
massive and therefore luminous stellar content. This is consistent with the
large column densities ($\ga10^{24}$~\cmq) derived in previous studies (see
Eq.~3 of Beltr\'an et al.~\cite{belt18}). The overall picture is suggestive
of a champagne flow confined to the east and expanding westwards. We
present additional evidence for this interpretation in Sect.~\ref{skin}.

\subsection{Recombination lines}
\label{srecl}

\begin{table}
\centering
\caption[]{
Hydrogen recombination lines observed in the present study.
}
\label{trec}
\begin{tabular}{cc|cc}
\hline
\hline
Transition & $\nu$ (MHz) &
Transition & $\nu$ (MHz) \\
\hline
H38$\alpha$$^a$ & 115274.399 &
H52$\beta$ & 88405.687 \\
H39$\alpha$ & 106737.357 &
H54$\gamma$ & 115570.116 \\
H40$\alpha$ & 99022.952 &
H55$\gamma$ & 109536.001 \\
H41$\alpha$ & 92034.434 &
H56$\gamma$ & 103914.838 \\
H42$\alpha$ & 85688.39 &
H57$\gamma$ & 98671.891 \\
H48$\beta$ & 111885.070 &
H58$\gamma$ & 93775.871 \\
H49$\beta$ & 105301.857 &
H59$\gamma$$^b$ & 89198.545 \\
H50$\beta$ & 99225.208 &
H60$\gamma$ & 84914.394 \\
H51$\beta$ & 93607.316 & \\
\hline
\end{tabular}

\vspace*{1mm}
$^a$~not used because heavily blended with the \CO(1--0) line \\
$^b$~not used because affected by \HCOp(1--0) in absorption
\end{table}

\begin{figure*}
\centering
\resizebox{15cm}{!}{\includegraphics[angle=0]{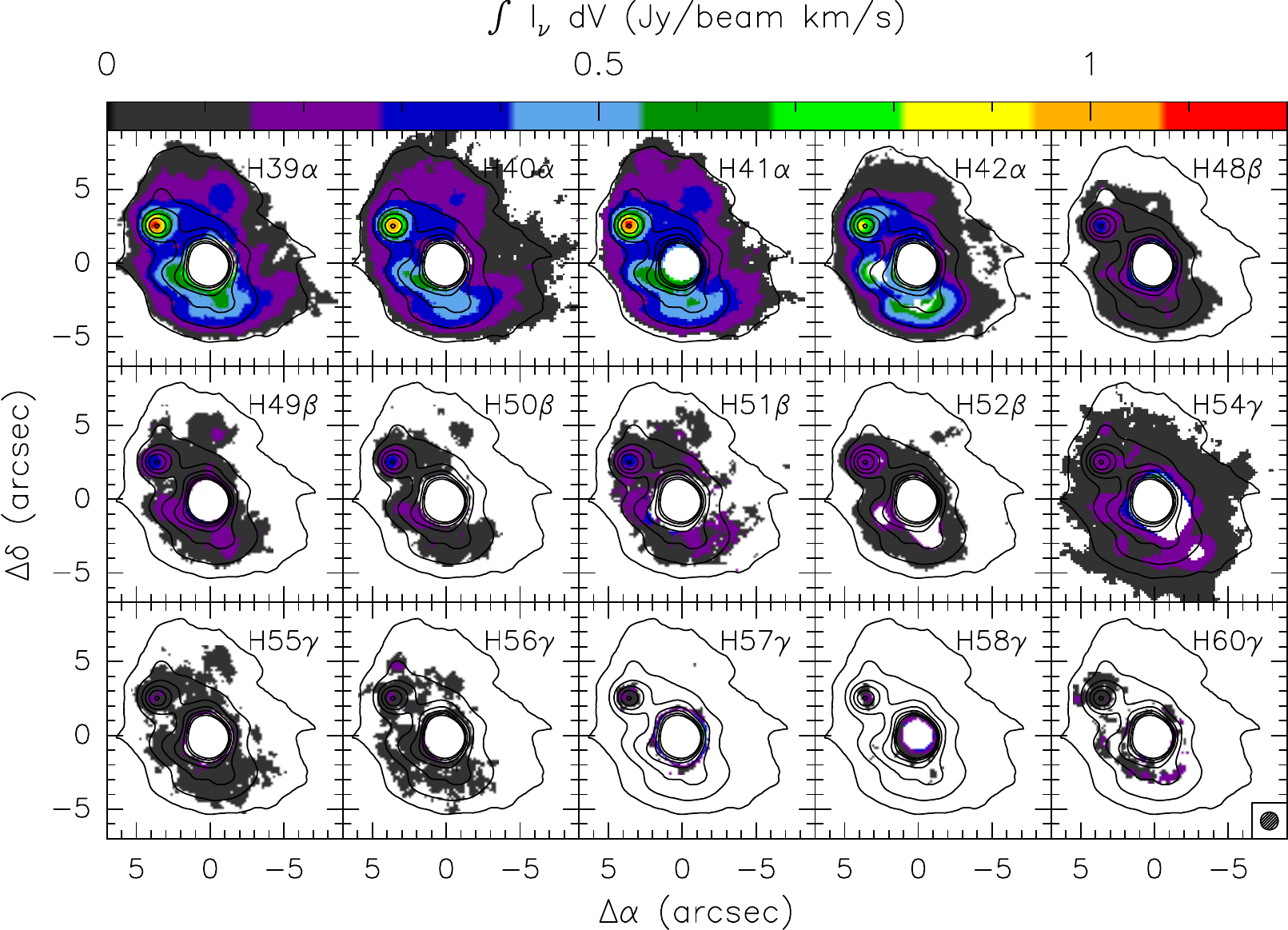}}
\caption{
Maps of the integrated emission over the hydrogen recombination lines
indicated in the top right corner of each panel. The emission from the HMC
is not plotted because the recombination lines are heavily blended with
stronger molecular transitions. The circle in the bottom right corner is
the synthesised beam. The offsets are relative to the phase center of the
ALMA observations. The emission of the H54$\gamma$ line is contaminated
by two hyperfine components of the NS~$J=5/2\rightarrow3/2$ transition.
}
\label{frint}
\end{figure*}

\begin{figure*}
\centering
\resizebox{15cm}{!}{\includegraphics[angle=0]{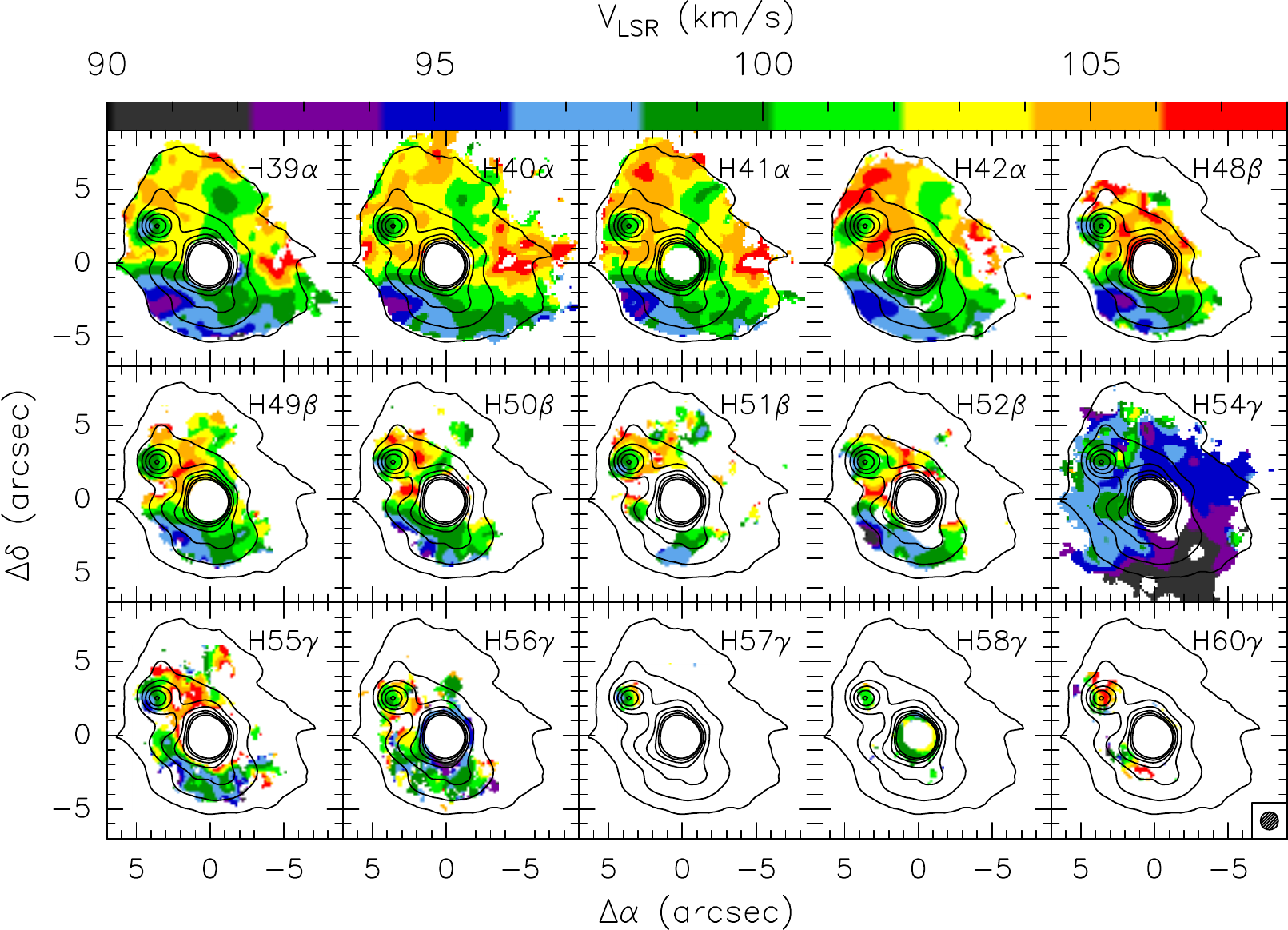}}
\caption{
Same as Fig.~\ref{frint}, for the peak velocity of the recombination lines.
}
\label{frvlsr}
\end{figure*}

\begin{figure*}
\centering
\resizebox{15cm}{!}{\includegraphics[angle=0]{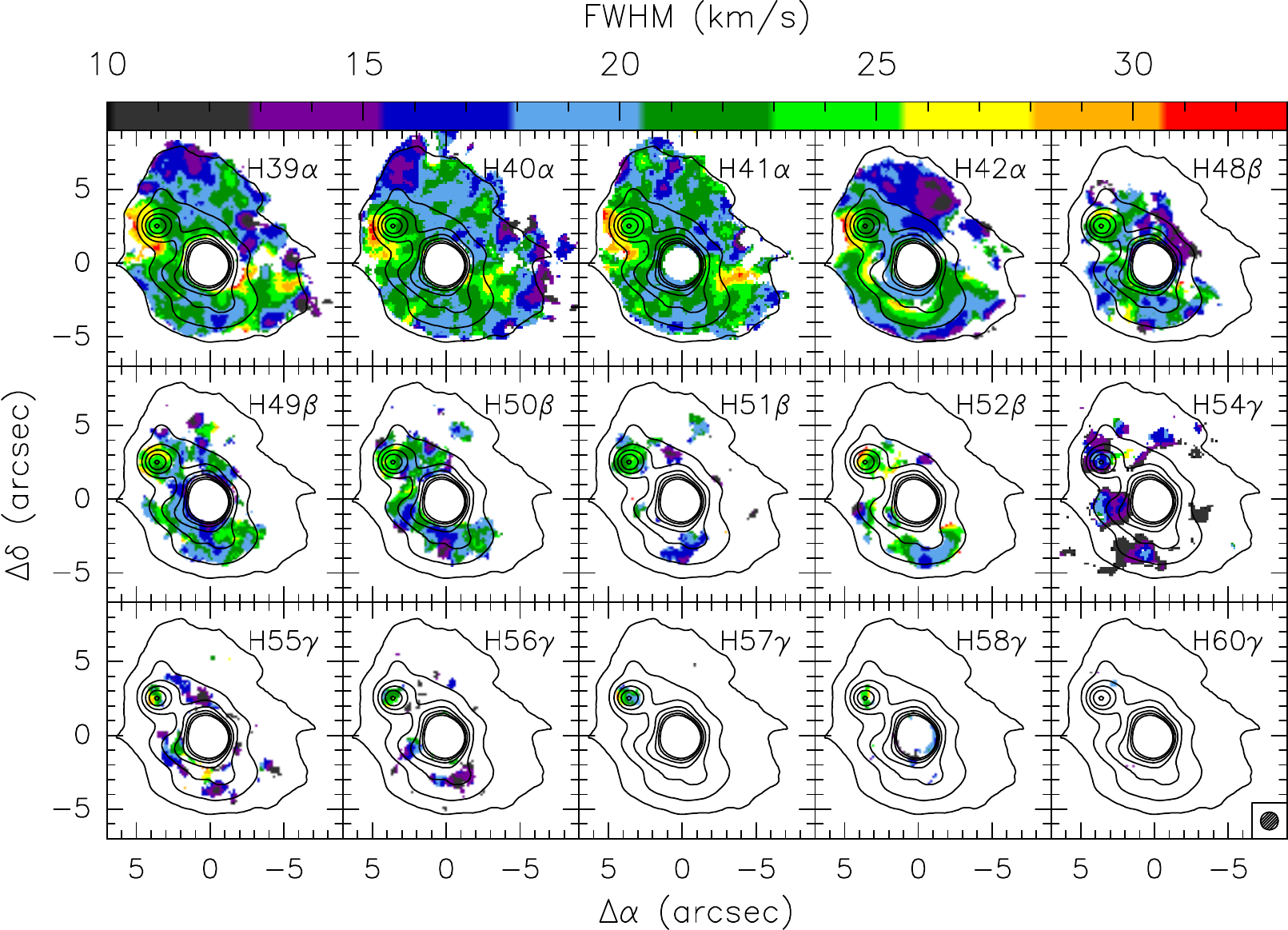}}
\caption{
Same as Fig.~\ref{frint}, for the full width at half maximum of the
recombination lines.
}
\label{frfwhm}
\end{figure*}

Thanks to the broad frequency coverage of the GUAPOS project, we could
reveal all of the $\alpha$, $\beta$, and $\gamma$ recombination lines
of hydrogen that fall in the ALMA 3-mm receiver band, with the exception
of the H38$\alpha$ line, which is heavily blended with the \CO(1--0)
transition, and the H59$\gamma$ line, which is faint and contaminated by an
\HCOp(1--0) absorption feature (see Table~\ref{trec}). No recombination
line of elements heavier than hydrogen was detected. In Figs.~\ref{frint},
\ref{frvlsr}, and~\ref{frfwhm} we show maps of the integrated intensity,
peak velocity, and full width at half maximum (FWHM) of the lines. These
parameters were obtained by fitting the spectrum at each pixel of the
map with a Gaussian. In this procedure, we rejected the data if less
than 5 channels were above 3$\sigma$ or the RMS of the residual spectrum
(after subtracting the fit) was above three times the noise of the original
spectrum. The latter criterion explains why the area corresponding to the
HMC is blanked in the maps, as the corresponding spectra present a line
forest that introduces a large RMS in the residual spectra.

It is evident that only the $\alpha$ lines trace the whole ionised
halo, whereas the $\gamma$ lines are detected mostly towards the \UC\
region.
This difference is not surprising as the latter lines are intrinsically
fainter than the former.
We note that the H54$\gamma$ line emission is contaminated by
the NS $J$=5/2--3/2 $\Omega$=1/2, $F$=5/2--3/2 $l$=$f$ and $F$=3/2--1/2
$l$=$f$ hyperfine components. The velocity maps outline a clear gradient
with velocity increasing from SSE to NNW. The \UC\ region is somehow
part of the general trend as one sees a velocity gradient across it in
the E--W direction, approximately the same as the symmetry axis of the
cometary shaped free-free emission (see Sect.~\ref{shii}). If we refer
the recombination line velocities to the systemic LSR velocity of the HMC
(96.5~\kms; see e.g. Cesaroni et al.~\cite{cesa11}), we see that most of the
ionised gas is red shifted, which implies that the \HII\ region is expanding
away from the dense molecular gas. Such a scenario is consistent with the broad
line width observed at the eastern side of the \UC\ region (see
Fig.~\ref{frfwhm}), which hints at a larger velocity dispersion at the
interface between the ionised gas and the dense molecular gas.

\subsection{Molecular lines}
\label{smolin}

\begin{figure}
\centering
\resizebox{8.5cm}{!}{\includegraphics[angle=0]{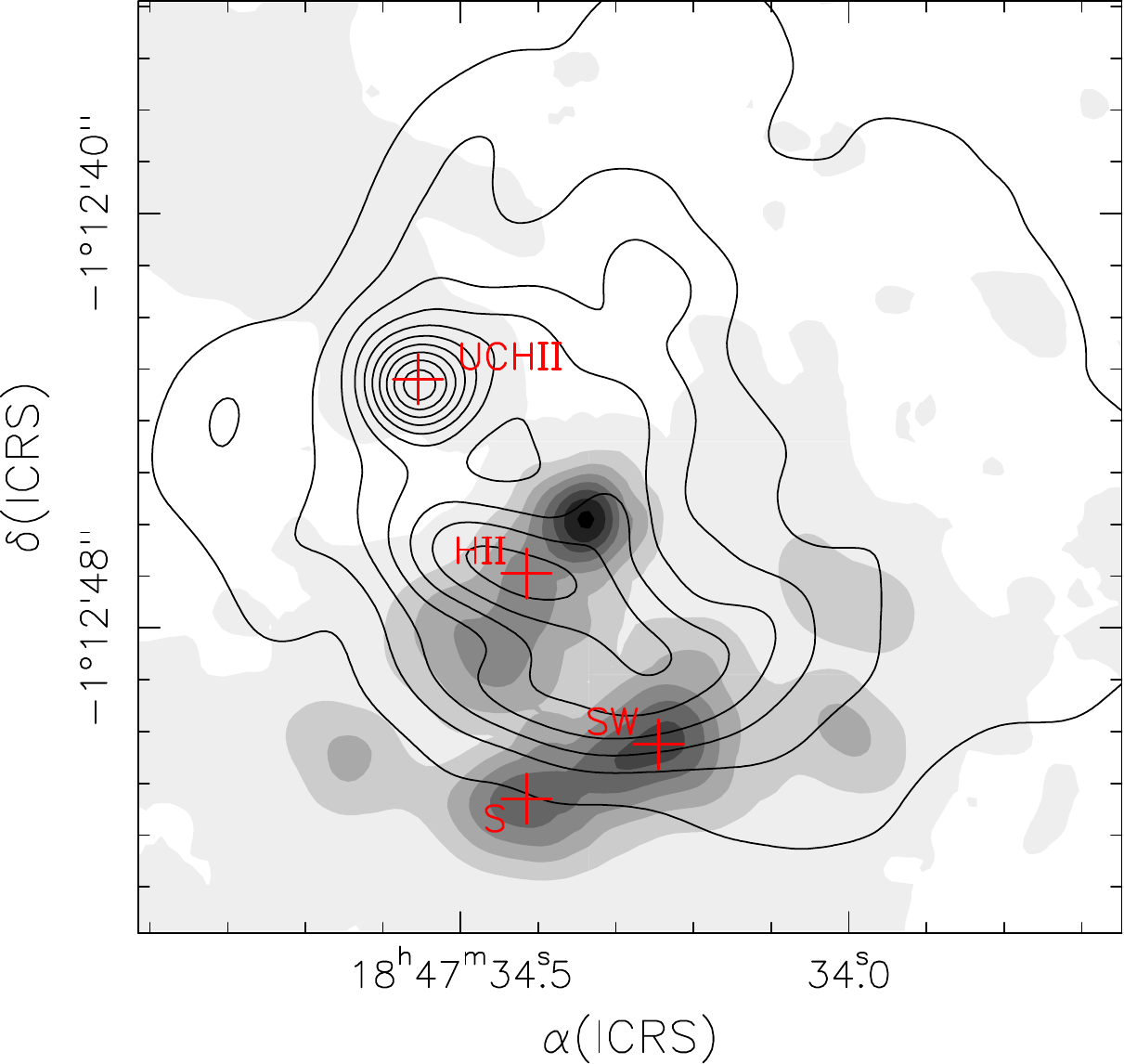}}
\caption{
Contour map of the 1~cm continuum emission overlaid on a map of the
peak emission in the \HCOpI(1--0) line (grey scale).  Contour levels range
from 1 to 28 in steps of 3~mJy/beam. The crosses and corresponding
labels denote the four template positions towards which the spectra of
Figs.~\ref{fspeca}--\ref{fspecd} have been taken.
}
\label{fpeaks}
\end{figure}

\begin{figure}
\centering
\resizebox{8.5cm}{!}{\includegraphics[angle=0]{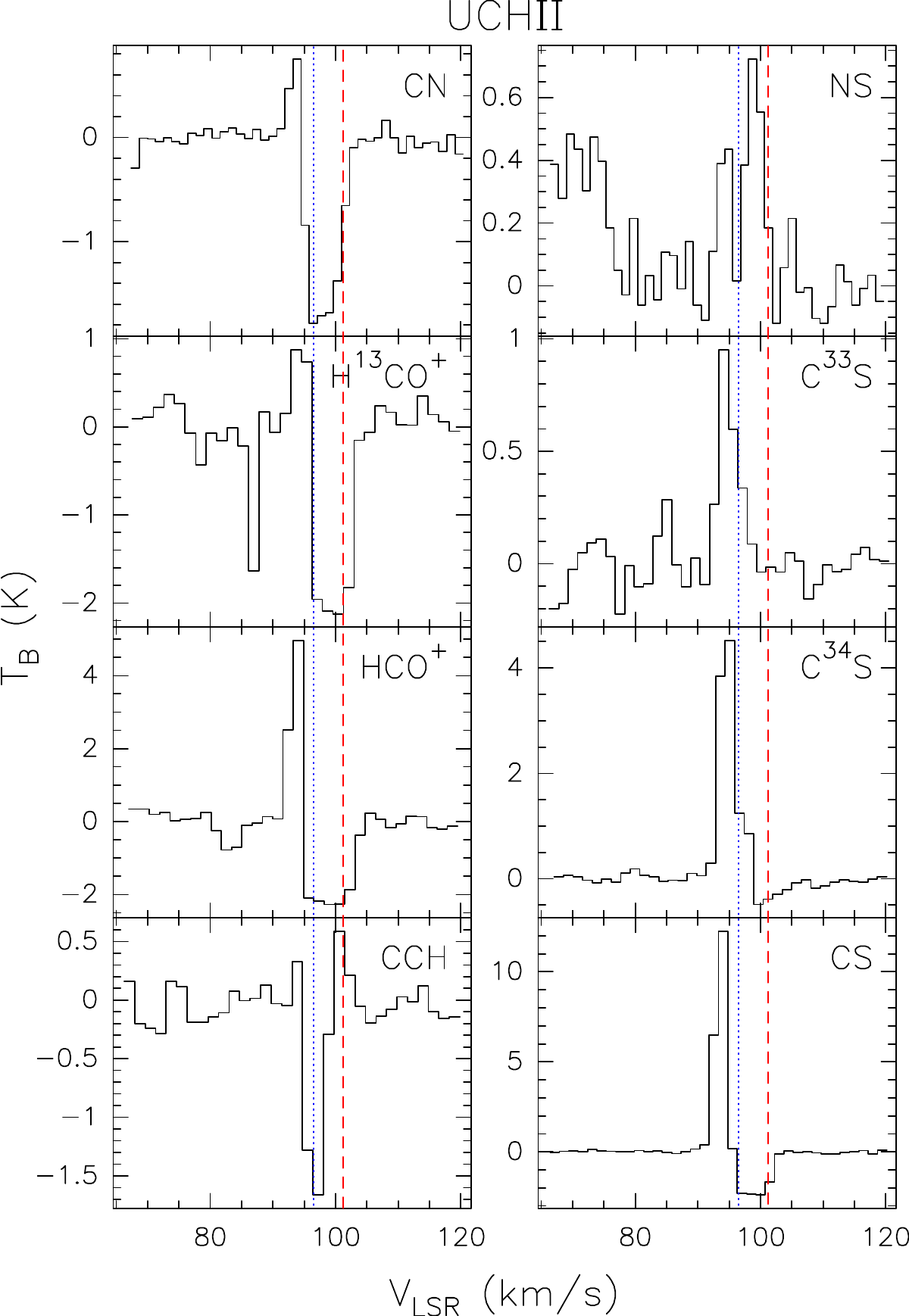}}
\caption{
Spectra of various molecular lines (as indicated in each panel) observed
towards the UCHII position in Fig.~\ref{fpeaks}
with ICRS coordinates $18^{\rm h}47^{\rm m}34\fs555$ --01\degr12\arcmin43\farcs20.
The blue dotted and red
dashed lines mark, respectively, the systemic LSR velocity of the HMC
(96.5~\kms; Cesaroni et al.~\cite{cesa11}) and the peak velocity of the
H39$\alpha$ recombination line measured at the given position.
}
\label{fspeca}
\end{figure}

\begin{figure}
\centering
\resizebox{8.5cm}{!}{\includegraphics[angle=0]{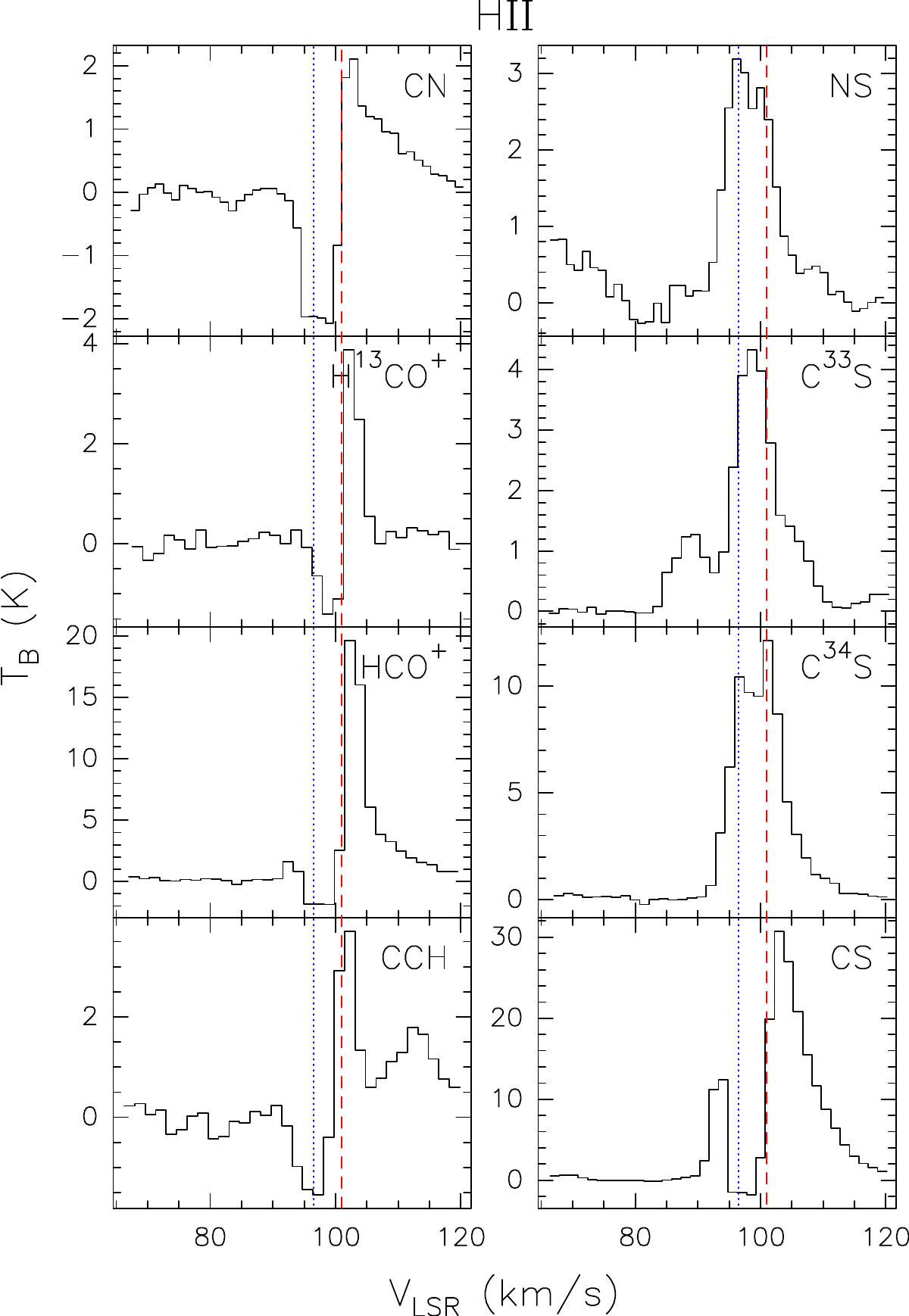}}
\caption{
Same as Fig.~\ref{fspeca}, for the HII position in Fig.~\ref{fpeaks}
with ICRS coordinates $18^{\rm h}47^{\rm m}34\fs415$ --01\degr12\arcmin46\farcs95.
}
\label{fspecb}
\end{figure}

\begin{figure}
\centering
\resizebox{8.5cm}{!}{\includegraphics[angle=0]{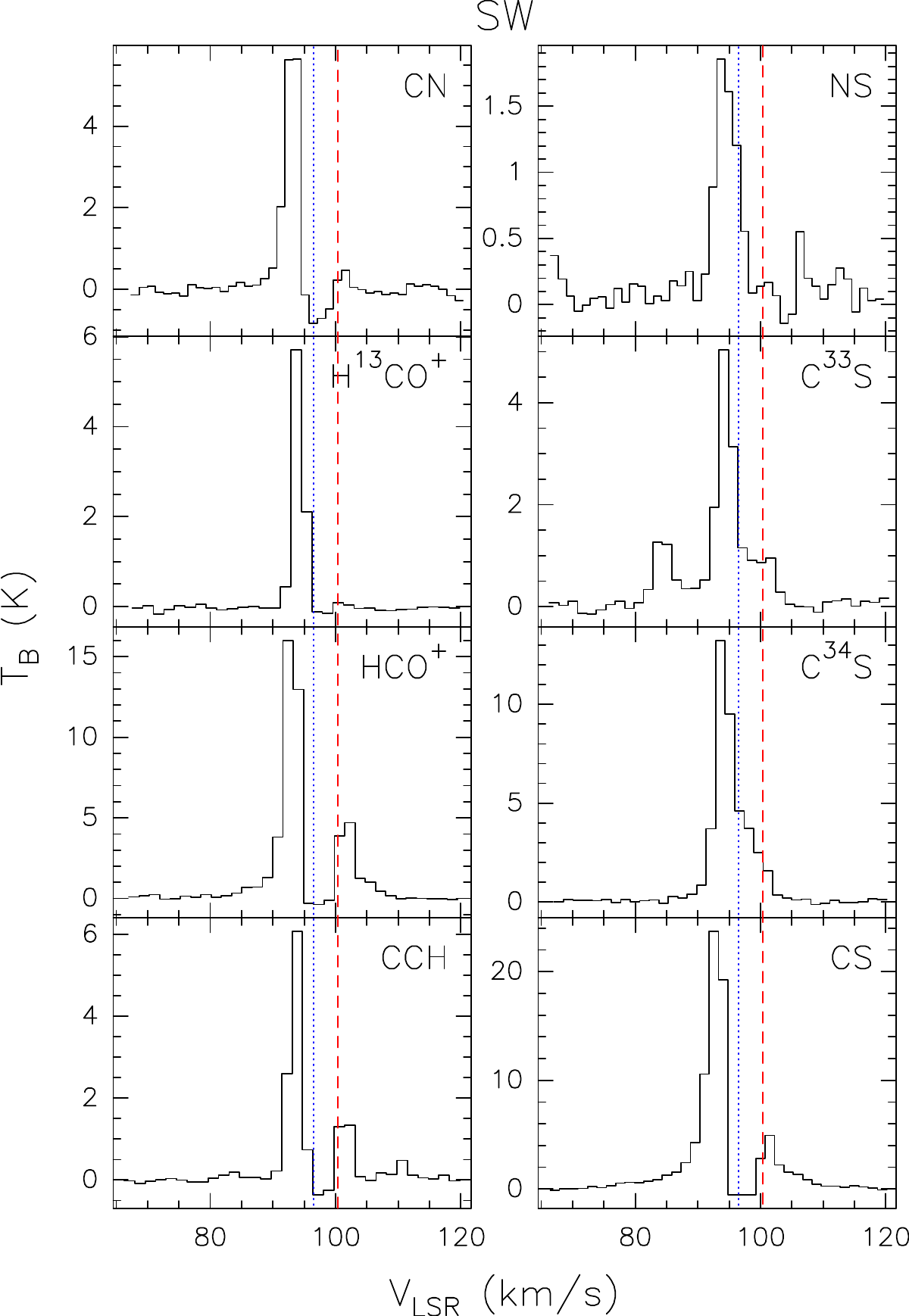}}
\caption{
Same as Fig.~\ref{fspeca}, for the SW position in Fig.~\ref{fpeaks}
with ICRS coordinates $18^{\rm h}47^{\rm m}34\fs245$ --01\degr12\arcmin50\farcs25.
}
\label{fspecc}
\end{figure}

\begin{figure}
\centering
\resizebox{8.5cm}{!}{\includegraphics[angle=0]{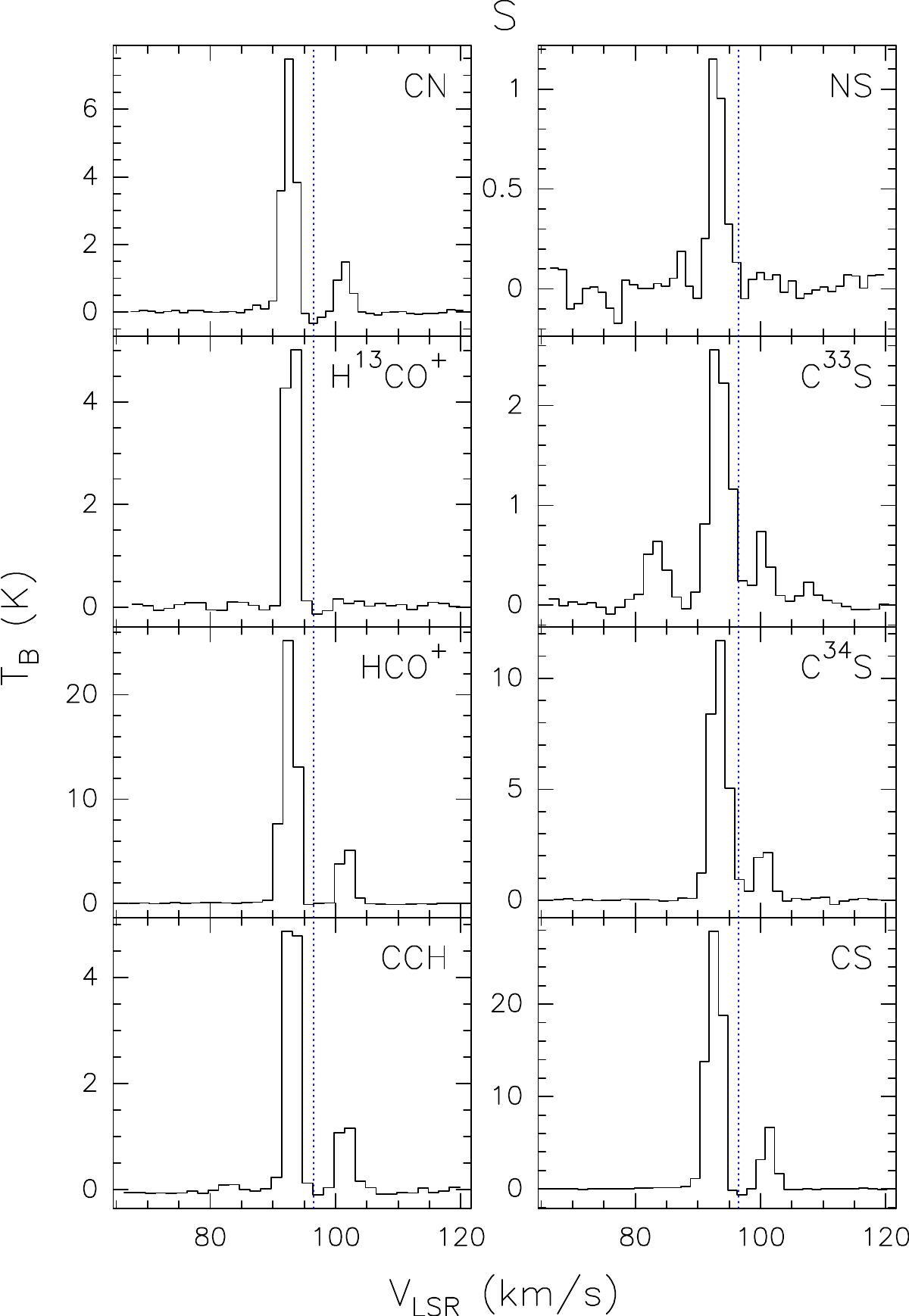}}
\caption{
Same as Fig.~\ref{fspeca}, for the S position in Fig.~\ref{fpeaks}
with ICRS coordinates $18^{\rm h}47^{\rm m}34\fs415$ --01\degr12\arcmin51\farcs30.
}
\label{fspecd}
\end{figure}

While the GUAPOS project is mostly focused on HMC tracers, the broad bandwidth
covers also other species that allow us
to study the more extended molecular clump enshrouding the \HII\
region. For this reason, we looked for rotational transitions of typical
medium-density ($\sim$$10^4$--$10^6$~\cmc) tracers. In particular, we selected the lines in
Table~\ref{tmol}, which were sufficiently intense and not affected by
blending with other lines. Figure~\ref{fpeaks} gives an idea
of the distribution of the medium-density molecular gas (traced by \HCOpI)
with respect to the \HII\ region and shows the four template positions
whose spectra are presented in Figs.~\ref{fspeca}--\ref{fspecd}. These
positions have been chosen at the main peaks of the free-free and \HCOpI(1--0)
line emission, excluding the HMC due to the complexity of the spectrum.
We note that the ``SW'' position lies very close to the ``shock region'' studied
by L\'opez-Gallifa et al.~(\cite{loga25}), also called ``region~2'' by
Fontani et al.~(\cite{font24}). The line profiles
present multiple velocity components in both emission and absorption.
It is worth noting that the absorption is seen only towards the bright
continuum background of the \HII\ region, which proves that it is real
and not a fake feature due to the presence of extended structures resolved
out by the interferometer.

\begin{table*}
\centering
\caption[]{
Parameters of the molecular lines selected for the study of the extended
molecular cloud enshrouding the \HII\ region.
}
\label{tmol}
\begin{tabular}{ccccc}
\hline
\hline
Molecule & Transition & $\nu$ (MHz) & $E_{\rm up}$ (K) & $S_{ij}\mu^2$ (D$^2$) \\
\hline
CN  & $N$=1--0, $J$=1/2--1/2, $F$=3/2--3/2 & 113191.2787 & 5.4 & 1.58359 \\
CCH  & $N$=1--0, $J$=3/2--1/2, $F$=2--1 & 87316.925 & 4.2 & 1.42458 \\
NS  & $J$=5/2--3/2, $\Omega$=1/2, $F$=7/2--5/2, $l$=$f$ & 115556.253 & 8.9 & 10.41507 \\
\HCOp  & $J$=1--0 & 89188.5247 & 4.3 & 15.21022 \\
\HCOpI  & $J$=1--0 & 89188.5247 & 4.3 & 15.21022 \\
CS  & $J$=2--1 & 97980.9533 & 7.1 & 7.64426 \\
C$^{34}$S  & $J$=2--1 & 96412.9495 & 7.6 & 6.94061 \\
C$^{33}$S  & $J$=2--1 & 97172.0639 & 7.0 & 30.60114 \\
CH$_3$CCH  & $J$=6--5, $K$=0,\dots,5 & 102499.0187--102547.9844 & 17.2--197.8 & 0.68608--2.24559 \\
\hline
\end{tabular}
\end{table*}

Although the complexity of the spectra makes the data interpretation more
difficult, the absorption features provide us with useful positional and
kinematical information on the neutral gas with respect to the ionised
gas. Looking at the spectra towards the UCHII and HII positions
(Figs.~\ref{fspeca} and~\ref{fspecb}), one sees that in both
positions absorption is detected in several lines at about the velocity of
the HMC, which suggests that the HMC itself may be part of a molecular clump
located between the \HII\ region and the observer. Another interesting
feature is the inverse P-Cygni profile seen towards the \UC\ region,
which might be due to residual infall. Towards the \HII\ region center
(Fig.~\ref{fspecb}) a broad red-shifted wing is present, which is related to
the red lobe of the most prominent bipolar outflow imaged by Beltr\'an et
al.~(\cite{belt18}). The absorption dip, slightly red-shifted with respect
to the HMC velocity, is likely due to the bright background continuum
emission of the nearby HMC. It is also worth noting that the blue-shifted
emission at $\sim$94~\kms\ seen at the S position (Fig.~\ref{fspecd}) could
be due to the molecular cloud imaged by Beltr\'an et al.~(\cite{belt22b})
to the south of the HMC. In Sects.~\ref{skin} and~\ref{sview} we analyse
the structure and kinematics of the region in more detail.

\section{Analysis}
\label{sana}

In this section we analyze the continuum and line data with the aim
to establish the physical properties of both the ionised and molecular
components and investigate the relationship between them.

\subsection{The \HII\ region}

From the integrated 1~cm continuum flux over the whole \HII\ region in
Fig.~\ref{fvmaps} one can compute the Lyman continuum photon rate, $N_{\rm
Ly}$, needed to ionise the gas. In this calculation we assume that the 1 cm
free-free emission is optically thin and the mean electron temperature is
6000~K (see Sect.~\ref{sfit}). We measure a flux density of $\sim$760~mJy
within the 3$\sigma$ contour level of the \HII\ region,
corresponding to $N_{\rm Ly}\simeq1.5\times10^{48}$~s$^{-1}$. If
the ionization is due to a single star, this value implies a
spectral type O8.5 (Panagia~\cite{pana}) and a stellar luminosity of
$\sim5\times10^4~L_\odot$. Cesaroni et al.~(\cite{cesa94a}) calculated
a bolometric luminosity from the IRAS fluxes of $6\times10^4~L_\odot$
(after scaling the value to the most recent estimate of 3.75~kpc),
only slightly greater (by $\sim10^4$~\Lsun, i.e. $\sim$20\%) than that obtained from the
Lyman continuum flux.

This result has the twofold implication that the ionizing flux must be
mostly due to a single massive star, and the number of Lyman continuum
photons absorbed by dust inside the \HII\ region is negligible. This is
because, on the one hand, for the same total bolometric luminosity multiple
early-type stars would emit less ionizing photons with respect to a single
star (see e.g. Fig.~7 of Cesaroni et al.~\cite{cesa15}); on the other
hand, (as noted by Wood \& Churchwell~\cite{wc89a}), if a significant
number of Lyman continuum photons were absorbed by dust inside the \HII\
region, the observed free-free emission should be fainter and suggest
a value of the stellar luminosity significantly less than $L_{\rm bol}$
(see Spitzer~\cite{spitzer}).

The previous conclusion -- namely that only one massive star contributes to the
bolometric luminosity and Lyman continuum -- is challenged by the findings
of Cesaroni et al.~(\cite{cesa10}) and Beltr\'an et al.~(\cite{belt21}) who
identify four massive young stellar objects (YSOs) inside the HMC. Being deeply
embedded, these (proto)stars cannot be responsible for the ionization of
the \HII\ region, but should contribute to the total luminosity with much more
than $\sim$$10^4$~\Lsun\ (see above). The solution of this conundrum might be
simply related to the uncertainties in the determination of the luminosities
and/or to the identification of the embedded YSOs. However, other explanations
are also possible. One is provided by the fact that $\sim$1/3 of the compact
\HII\ regions appear to be ionised by stars with Lyman continuum luminosities
more intense than expected for their spectral types -- the so-called
``Lyman excess'' (S\'anchez-Monge et al.~\cite{samo13}; Urquhart et
al.~\cite{urqu13}; Cesaroni et al.~\cite{cesa15,cesa16}).
If the star ionizing the \HII\ region belongs to this type of
objects, its contribution to the bolometric luminosity is less than the
$5\times10^4$~\Lsun\ previously estimated. The other possibility is that
part of the stellar photons longward of 912~\AA\ are not absorbed by the
dust and are leaking from the clump enshrouding the \HII\ region. This would
cause an underestimate of the bolometric luminosity because the latter is
computed from the IR emission of the clump. Indeed, the cometary
structure of the region is consistent with lower density, and thus lower
opacity, to the NW. A similar situation has been recently proposed for
Sgr~B2 (see Budaiev et al.~\cite{buda25}), where part of the radiation
appears to be escaping even from the densest regions.

\subsubsection{The model fit}
\label{smod}

\begin{figure*}
\centering
\resizebox{15cm}{!}{\includegraphics[angle=0]{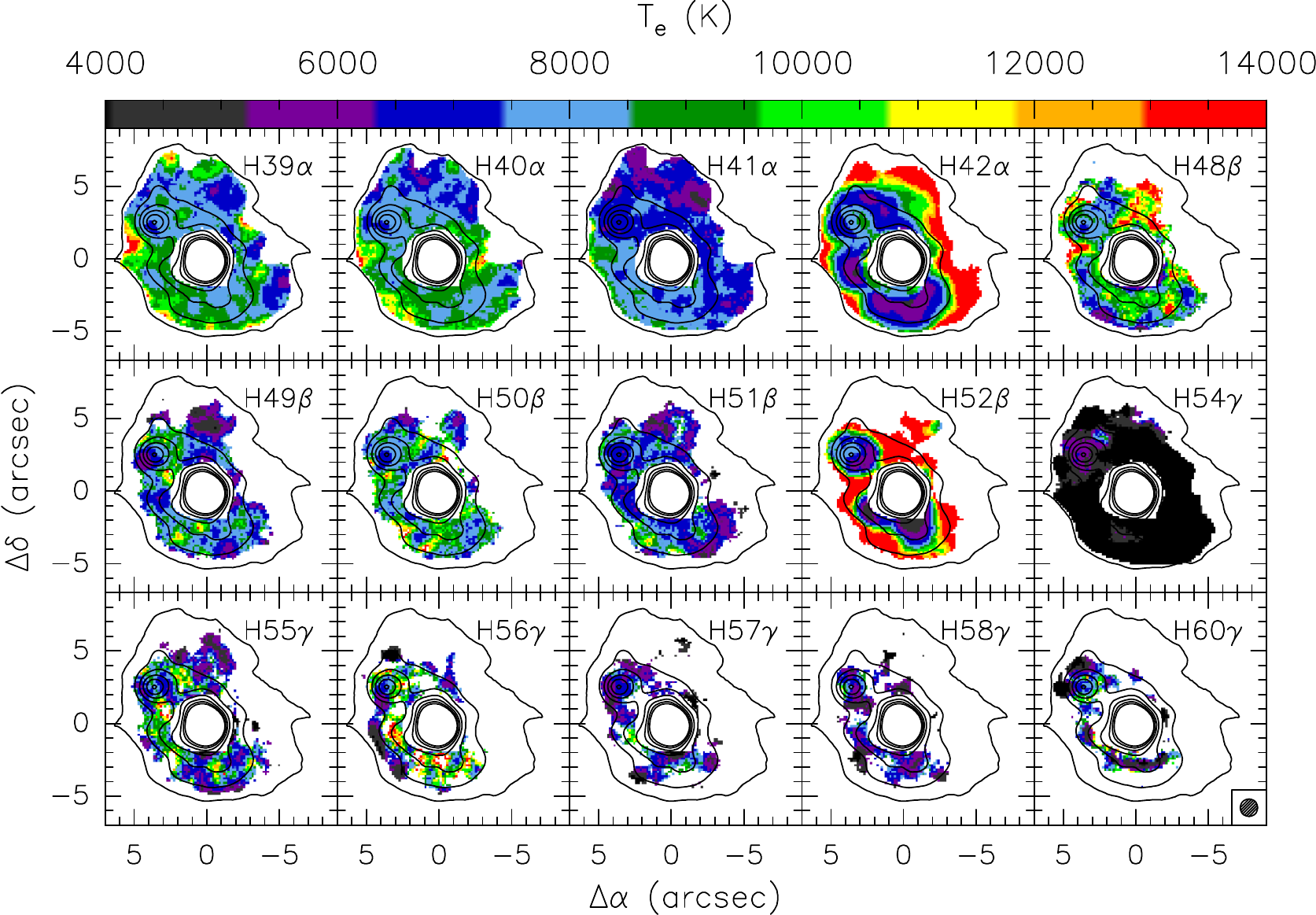}}
\caption{
Maps of the electron temperature obtained from the hydrogen recombination
lines indicated in the top right corners of the panels, under the LTE
approximation. The offsets are relative to the phase center of the ALMA
observations. The contours are the map of the 3~mm continuum emission shown
in Fig.~\ref{fvmaps}c, with contour levels ranging from 1.15 to 28.75 in steps
of 4.6~mJy/beam. The synthesised beam is shown in the bottom right corner.
}
\label{flte}
\end{figure*}

The detection of many recombination lines provides us with the opportunity
to produce temperature and density maps of the \HII\ region. An
approximate estimate of the electron temperature can be computed from
the line-to-continuum ratio using Eq.~(2.124) of Gordon \& Sorochenko
(\cite{goso}),
which assumes LTE.
By applying this expression to each line in each pixel of the
map, we obtain the electron temperature maps in Fig.~\ref{flte}. Although
the maps with the best signal-to-noise ratio appear roughly consistent
with each other, suggesting a value of $\sim$7000-8000~K, other maps
look significantly different.

Such a discrepancy could be partly due to the fact
that our calculation made use of the continuum map at the same frequency as
the line map. The millimeter continuum emission is not pure free-free but is
known to be contaminated by thermal dust emission (see Sect.~\ref{scont}).
To get rid of this problem, we employed the following procedure. For
each pixel, we subtracted a constant baseline (0th order polynomial)
from the spectrum, thus removing the continuum emission. Then we added
an ``artificial'' continuum level by extrapolating to the corresponding
frequency the VLA 1~cm continuum map in Fig.~\ref{fvmaps}, assuming
optically thin free-free emission (i.e. flux $\propto\nu^{-0.1}$). Since
the dust emission is negligible at centimeter wavelengths, the continuum
level created in this way should correspond to the pure free-free emission
emitted at the frequency of the recombination line.

Another explanation for the differences among the maps in Fig.~\ref{flte}
is that non-LTE effects could play an important role. To take these into
account,
the line+continuum spectra were fitted with the
model described in Cesaroni et al.~(\cite{cesa19}).
which calculates the intensity of the line+continuum emission for given
values of the electron density, \ne, and temperature, \Te.  Besides
the standard expression for the free-free opacity (e.g. Eq.~A.1a of
Mezger et al.~\cite{mezg67}), this model uses Eqs.~(3.19) and (3.25)
of Brocklehurst \& Seaton~(\cite{brse}) to calculate the emission and
absorption coefficients of the recombination line. The $b$ and $\beta$
coefficients expressing the departure from LTE are computed with the
code of Gordon \& Sorochenko~(\cite{goso}). Additional details can
be found in Cesaroni et al.~(\cite{cesa19}), who
assume an expanding shell \HII\
region with constant electron density \ne\ and temperature \Te. In our
application, we neglect the expansion and assume zero inner radius. The
model requires also knowledge of the source distance (3.75~kpc) and \HII\
region radius. In our case, the \HII\ region is far from an ideal uniform
sphere and we are thus forced to adopt an approximation to estimate a
reasonable value of the radius.  Appendix~\ref{arad} describes how we
calculate the radius, which ranges between 3\farcs8 ($\sim$0.07~pc) and
7\farcs6 ($\sim$0.14~pc).

It is also necessary to have a reasonable guess of the turbulent velocity, $\Vt$,
contributing to the Doppler line width as in Eq.~(1) of Gordon \& Churchwell~(\cite{goch70}):
\begin{equation}
 \Delta V_{\rm D}^2 = 4 \ln2 \left( \frac{2kT_{\rm e}}{m} + \frac{2}{3}\Vt^2 \right)   \label{edw}
\end{equation}
where $k$ is the Boltzmann constant and $m$ the mass of the atom.
This parameter can be derived if two measurements of the line width
are available for the same recombination line of two different atomic
species. Unfortunately, only hydrogen lines have been detected in our
observations of \G, but we have available single-dish spectra of the
55$\alpha$, 57$\alpha$, and 58$\alpha$ recombination lines of both hydrogen and
helium, obtained with the Yebes 40-m telescope (V. Rivilla,
priv. comm.). From these, we obtained the line FWHM with Gaussian fits
and computed the weighted mean of the FWHM, which is 22.1$\pm$0.1~\kms\ for
H and 15.4$\pm$1.0~\kms\ for He. Then using Eqs.~(2) and~(3) of Gordon \&
Churchwell~(\cite{goch70}) we derived mean values over the whole \HII\
region (unresolved in the single-dish beam of 42\arcsec--54\arcsec)
of $\Vt$=9.1$\pm$1.4~\kms\ and \Te=7200$\pm$900~K, a temperature consistent
with the values in Fig.~\ref{flte}. For
our convenience we define a ``turbulent temperature'' $\Tt=m_{\rm
H}\Vt^2/(3k)=5030\pm1360$~K, with $m_{\rm H}$ mass of the H atom and $k$
Boltzmann constant. Hence, Eq.~(\ref{edw}) for the H atom takes the form
\begin{equation}
 \Delta V_{\rm D}^2 = 8 \ln2 ~ k \, \frac{T_{\rm e}+\Tt}{m_{\rm H}}.
 \label{ett}
\end{equation} In conclusion, since \Te$\simeq$$\Tt$ within the errors,
in our model we assumed $\Tt$=\Te\ all over the \HII\ region.

To get rid of the systemic velocity of the ionised gas and allow comparison
with the model, for each spectrum we subtracted the peak velocity, obtained
with a Gaussian fit as explained in Sect.~\ref{srecl}.
Then we convolved the model brightness temperature of the line+continuum emission
with the synthesised beam of the observations (1\farcs2) and computed the
$\chi^2$ at each position of the map with the expression
\begin{equation}
\chi^2 = \sum_i \frac{1}{\sigma_i^2} \sum_j \left( T_i^{\rm m}(V_j) - T_i^{\rm d}(V_j) \right)^2
   \label{echiq}
\end{equation}
where $i$ indicates the recombination line, $j$ the spectral channel,
$\sigma_i$ the noise of the spectrum, $V_j$ the channel velocity,
$T_i^{\rm d}$ the observed brightness temperature of the line+continuum,
and $T_i^{\rm m}$ that obtained from the model after convolution with the
1\farcs2 beam. In this calculation we considered only the three strongest
recombination lines, namely H39$\alpha$, H40$\alpha$, and H41$\alpha$. The
best fit was obtained by minimizing $\chi^2$ after varying the only two
free parameters of the model, \ne\ and \Te, over suitable ranges. The
uncertainty on the best-fit parameters was estimated with the method of
Lampton et al.~(\cite{lamp76}).

\subsubsection{Results of the fit}
\label{sfit}

\begin{figure}
\centering
\resizebox{8.5cm}{!}{\includegraphics[angle=0]{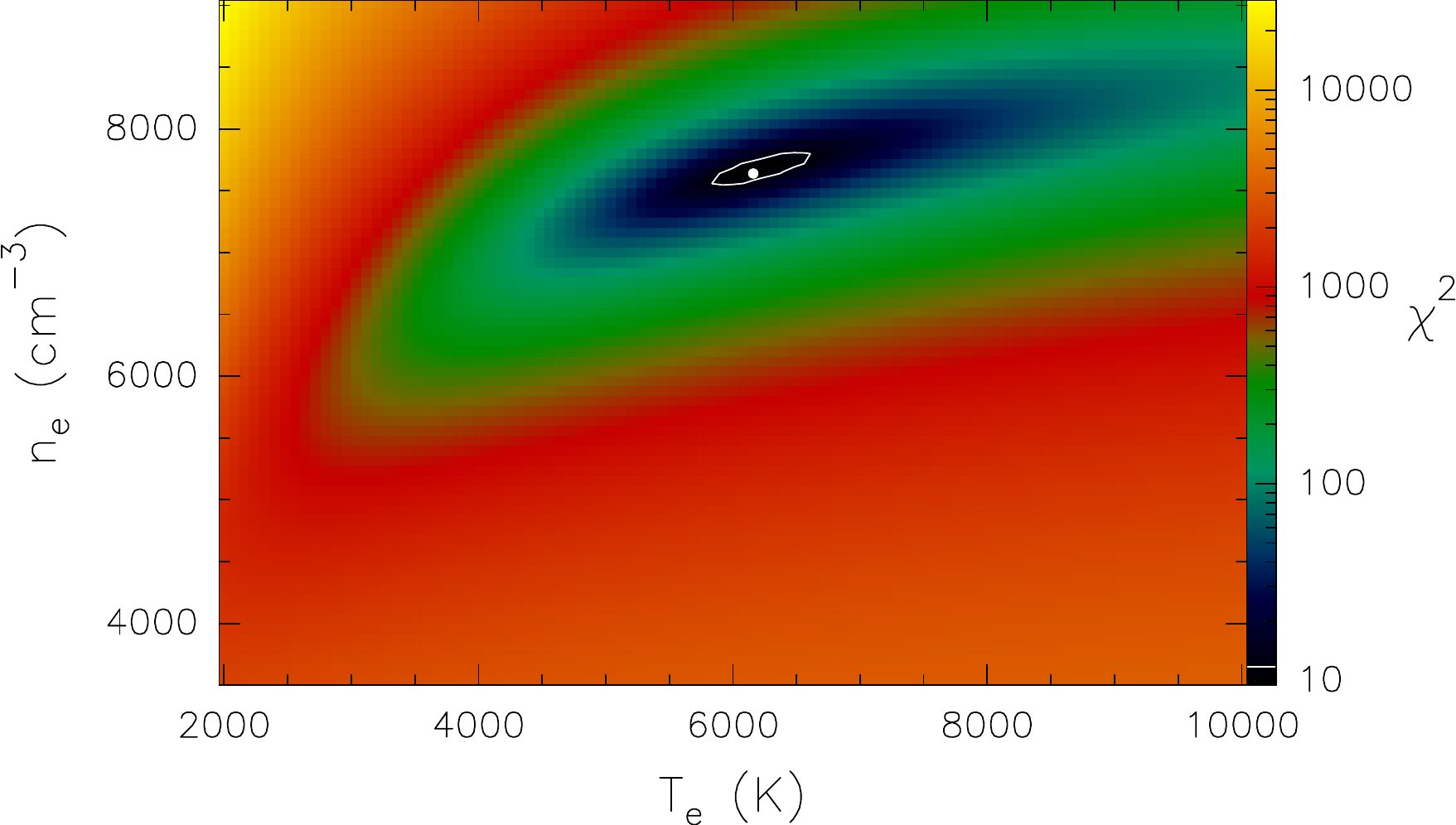}}
\caption{
Plot of the $\chi^2$ computed from Eq.~(\ref{echiq}) as a function of electron temperature
and density at the peak position of the \UC\ region. The white dot marks the minimum and
the white contour corresponds to the 1$\sigma$ confidence level.
}
\label{fchiq}
\end{figure}

\begin{figure}
\centering
\resizebox{8.5cm}{!}{\includegraphics[angle=0]{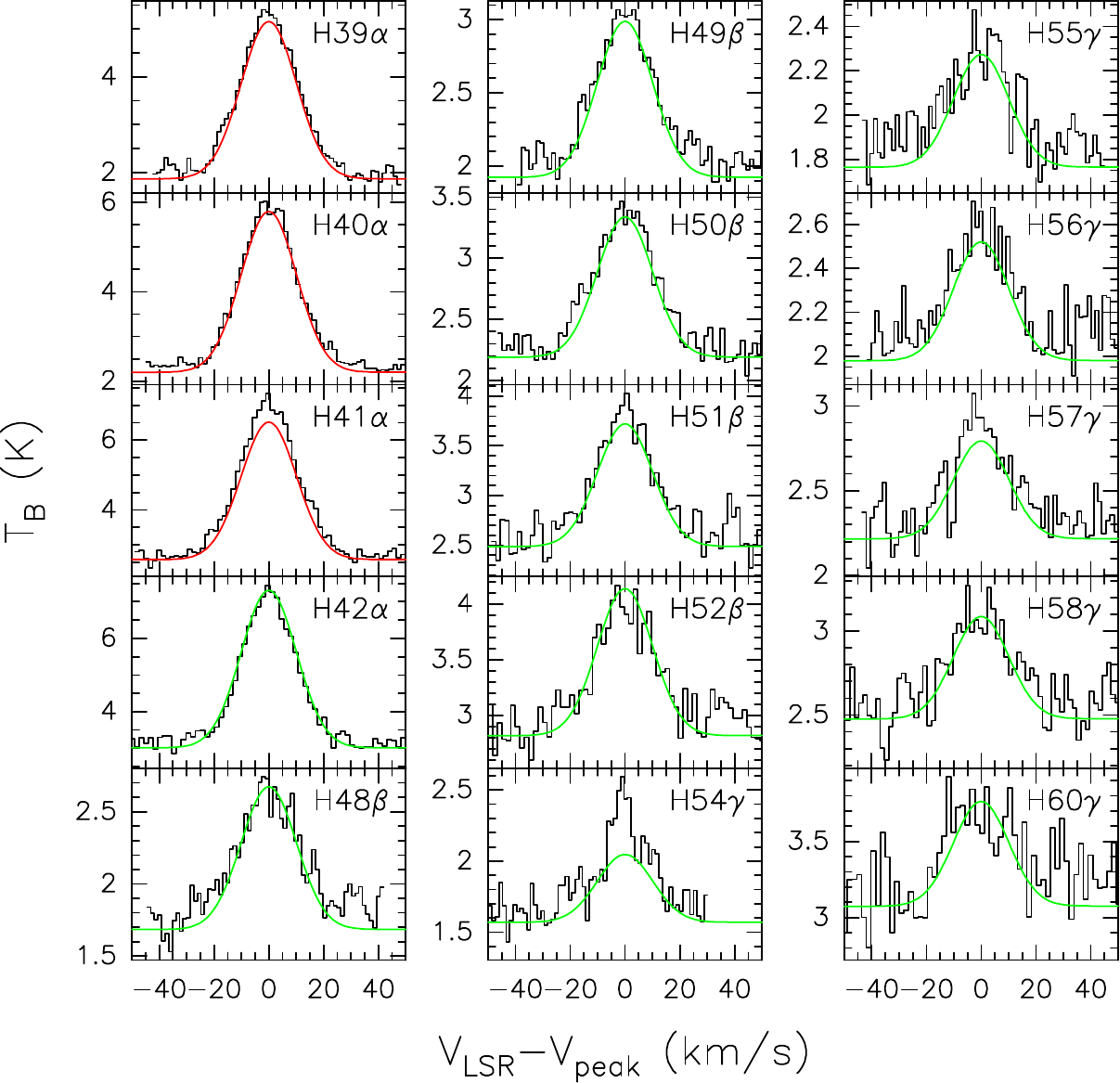}}
\caption{
Spectra of the hydrogen recombination lines observed towards the peak of
the \UC\ region. The name of the line is indicated in the top right corner
of each panel. The velocity is relative to the peak velocity of
each spectrum, obtained with a Gaussian fit. The red and green curves are
the model spectra obtained by fitting only the H39$\alpha$, H40$\alpha$,
and H41$\alpha$ lines. The H54$\gamma$ line is contaminated by emission
from a molecular line.
}
\label{flfit}
\end{figure}

An example of the $\chi^2$ minimization is illustrated in
Fig.~\ref{fchiq} for the peak position of the \UC\ region, which gives
\ne=$7640^{+168}_{-92}$~cm$^{-3}$ and \Te=$6160^{+450}_{-325}$~K.  In the
first three panels of Fig.~\ref{flfit} we show the corresponding best-fit
spectra (red curves) overlaid on the observed ones. In the other panels
of the same figure we also show the model spectra (green curves) for the
other recombination lines, computed using the same best-fit parameters. We
stress that, although the fit was obtained using only three lines, the
match is excellent also for all the others\footnote{As already noted in
Sect.~\ref{srecl}, the H54$\gamma$ line partly overlaps with two
of the NS $J$=5/2$\rightarrow$3/2 hyperfine components.}, which
demonstrates the reliability of our method.

\begin{figure}
\centering
\resizebox{8.5cm}{!}{\includegraphics[angle=0]{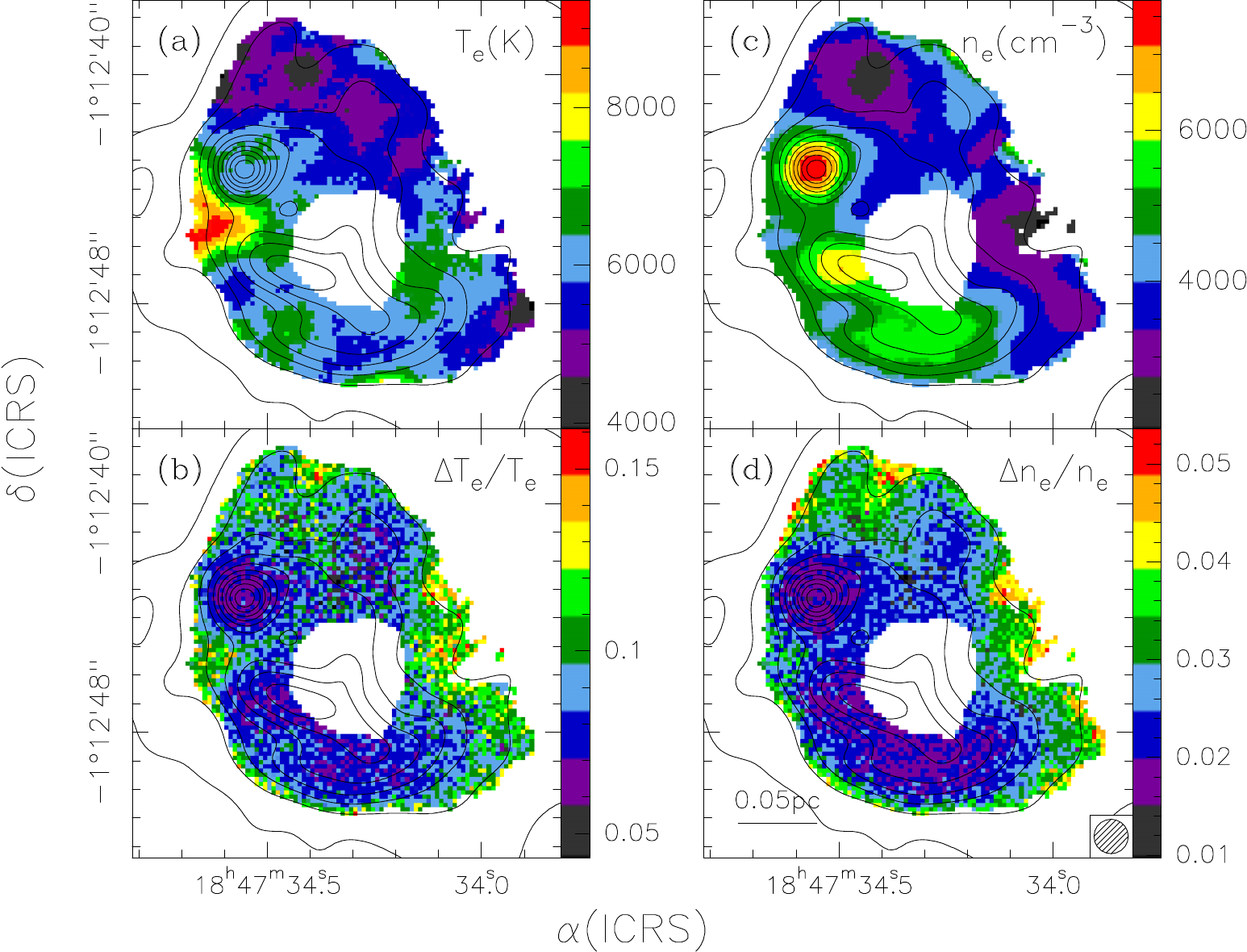}}
\caption{
{\bf a.} Map of electron temperature of the \HII\ region, obtained by
fitting the hydrogen recombination lines with our model (see text).
{\bf b.} Map of the relative error on the electron temperature.
{\bf c.} Same as panel a, for the electron number density.
{\bf d.} Same as panel b, for the electron number density.
The angular resolution is represented by the circle in the bottom right corner.
}
\label{fnete}
\end{figure}

The maps of \ne\ and \Te\ in Fig.~\ref{fnete} were obtained by fitting the
data only in those pixels where all three fitted lines where clearly
detected. The criterion for a detection is that at least 5 line channels
must have intensities above $3\sigma$. We note that all pixels corresponding
to the HMC (blanked in Figs.~\ref{frint}--\ref{frfwhm}) were
rejected because of heavy overlap with molecular lines,
which fill almost the whole observed bandwidth (see Fig.~4 of Mininni et
al.~\cite{mini20}).
The resulting
values are affected by relatively small uncertainties, less than 15\%
for \Te\ and less than 5\% for \ne.

The electron temperature is roughly constant all over the
\HII\ region, with a mean value of $\sim$6000~K
and a median of $\sim$4600~K, comparable to the
estimate of 5000~K obtained by Wood \& Churchwell~(\cite{wc91}) in a similar
object, the cometary \UC\ region G29.96--0.02. The prominent peak of
$\sim$9000~K close to the SE border of the \UC\ region might be real, but
it is also possible that it is due to an enhancement of the
turbulent velocity at the interface between the \UC\ region and the dense
molecular gas located to the east. In this case, our assumption of constant
$\Tt$=\Te\ could underestimate the contribution of $\Vt$ to the line width,
thus overestimating the value of \Te\ needed to fit the line profile.
The electron density distribution basically mirrors the intensity of the
free-free emission, with a main peak toward the \UC\ region and a secondary
peak between the HMC and the molecular gas located to the south (see
Sect.~\ref{smolc}).

\subsection{The molecular clump}
\label{smolc}

The continuum and molecular line maps of the GUAPOS project can be used to
obtain information on the dusty molecular gas enshrouding the \HII\ region.
In the following we derive estimates of several physical parameters of
this neutral component.

\subsubsection{Velocity, temperature, and column density from line emission}

\begin{figure}
\centering
\resizebox{7.5cm}{!}{\includegraphics[angle=0]{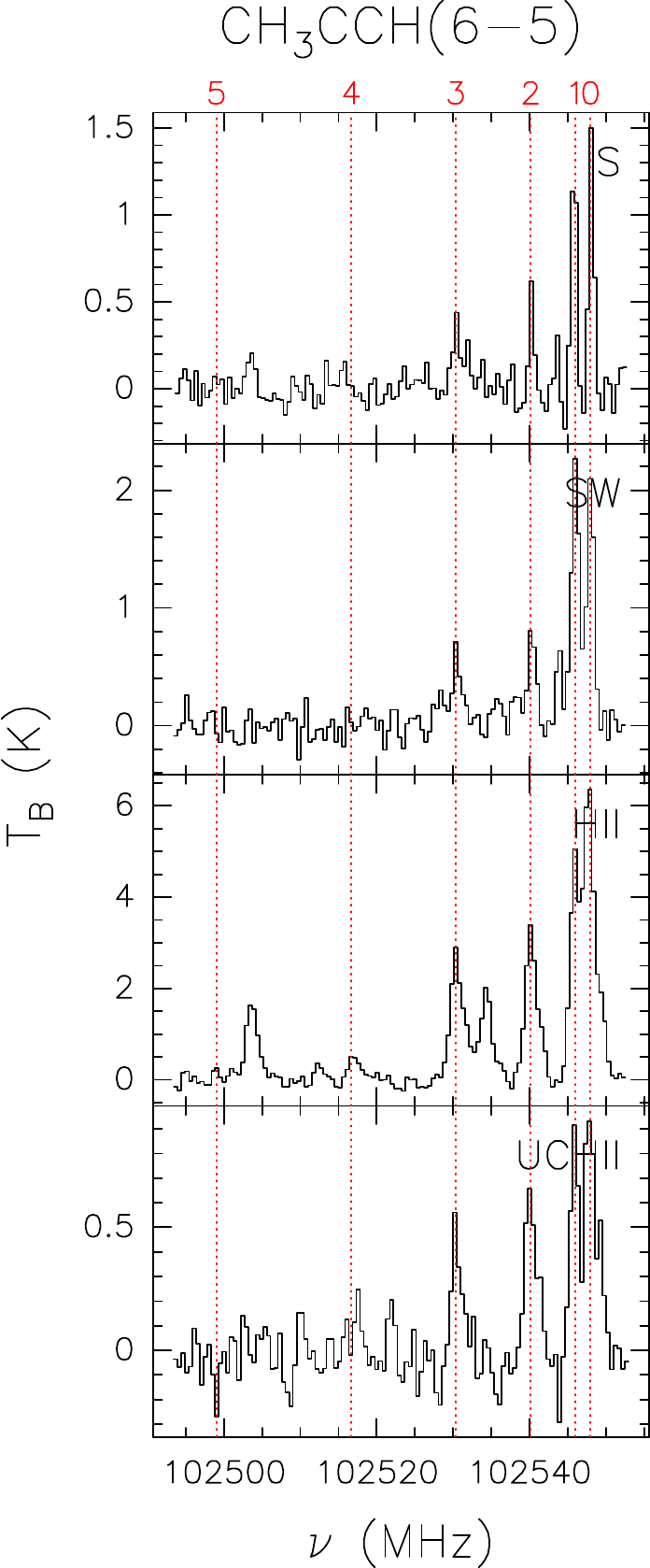}}
\caption{
Spectra of the \MAC(6--5) transition towards the four positions marked in
Fig~\ref{fpeaks}. The $K$=0 to 5 components are indicated by the red dotted
lines and labelled accordingly.
}
\label{fspmac}
\end{figure}

As discussed in Sect.~\ref{smolin}, even the molecular lines tracing
the extended, lower-density medium have complex spectra with multiple
components both in emission and absorption. Since we want to derive maps of
physical parameters, it is necessary to select the molecular transitions
that are least affected by this problem. After inspecting the most common
molecular species, we decided to focus on \CSII\ and \MAC, whose emission
is dominated by one velocity component, and CN, which shows a prominent
absorption feature. The \CSII\ molecule is especially useful to inspect the
velocity field of the dense material close to the \HII\ region surface,
while \MAC\ is a symmetric top molecule known to be an excellent tool to
derive the temperature and column density thanks to its multiple $K$-ladder
components (see e.g. Fontani et al.~\cite{font02}).
Template spectra of the \MAC(6--5) transition towards the four positions
marked in Fig.~\ref{fpeaks} are shown in Fig.~\ref{fspmac}.

\begin{figure}
\centering
\resizebox{8.5cm}{!}{\includegraphics[angle=0]{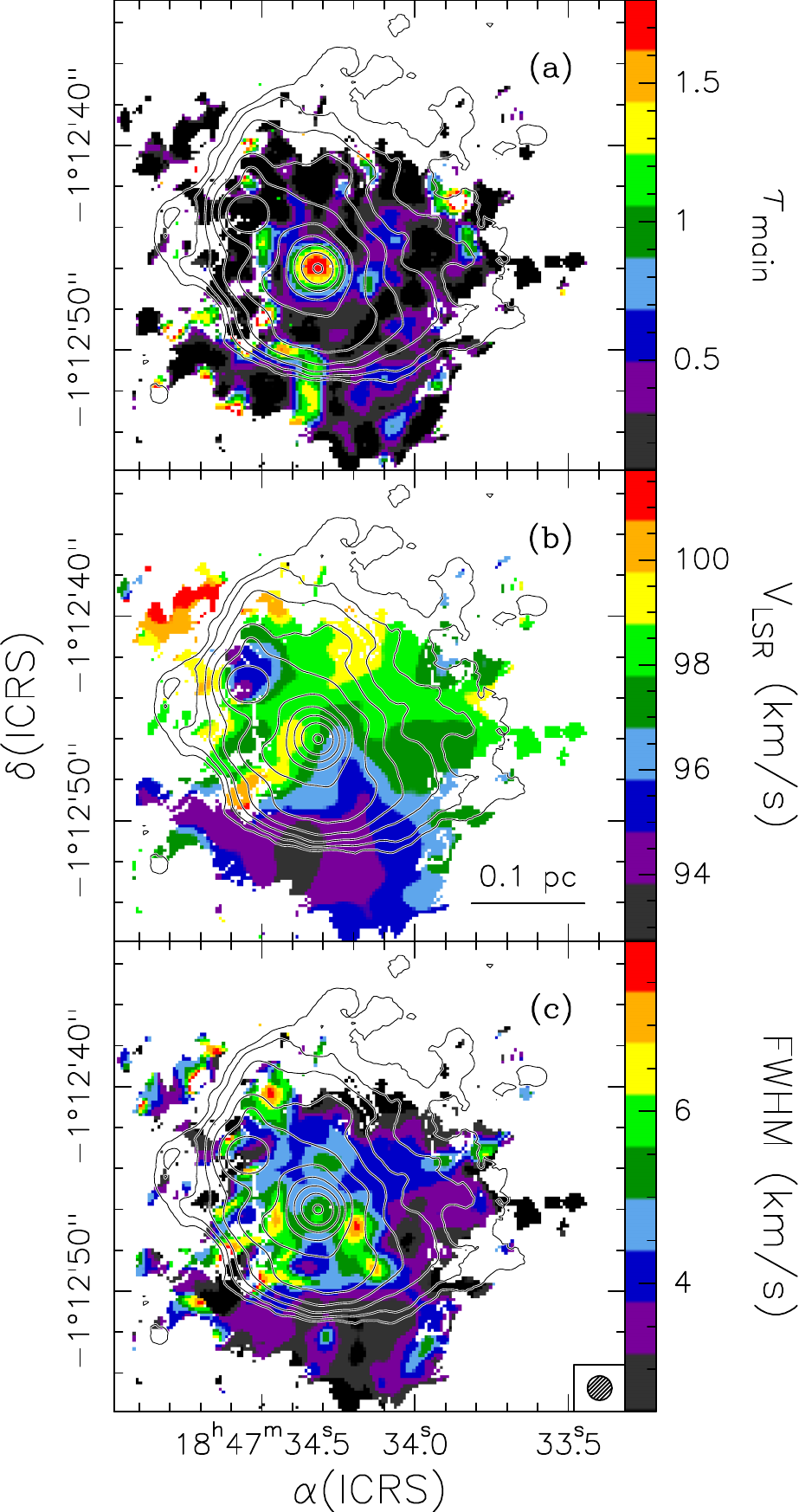}}
\caption{
{\bf a.} Map of the opacity of the main hyperfine transition of \CSII(2--1)
(colour image).  The contour map represents the 3~mm continuum emission
and is the same as in Fig.~\ref{fvmaps}c.
{\bf b.} Same as panel a, for the LSR velocity.
{\bf c.} Same as panel a, for the line FWHM.
The synthesised beam is shown in the bottom right corner.
}
\label{fctts}
\end{figure}

\begin{figure}
\centering
\resizebox{8.5cm}{!}{\includegraphics[angle=0]{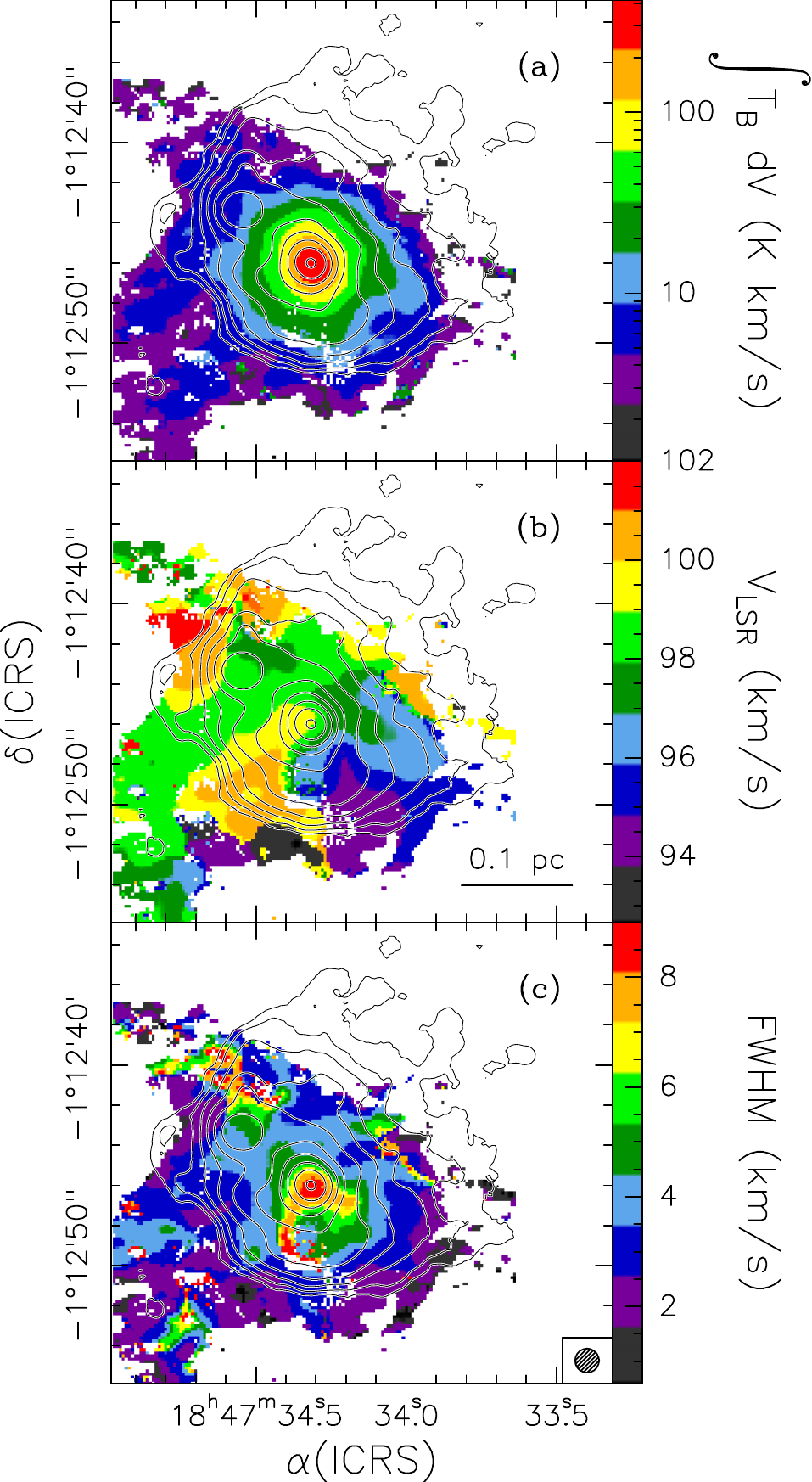}}
\caption{
{\bf a.} Map of the integrated intensity over the $K$=0 and 1 components
of the \MAC(6--5) line (colour image).  The contour map represents the
3~mm continuum emission and is the same as in Fig.~\ref{fvmaps}c.
{\bf b.} Map of the LSR velocity of the \MAC(6--5) line obtained by fitting
all the $K$ components (see Sect.~\ref{scont}).
{\bf c.} Same as panel b, for the line FWHM.
The synthesised beam is shown in the bottom right corner.
}
\label{fmac}
\end{figure}

We have fitted Gaussian profiles to the \CSII\ and \MAC\ spectra pixel by
pixel. It is worth noting that overlap with lines of other species may occur
only towards the HMC, given the chemical richness of this object (see
Fig.~4 of Mininni et al.~\cite{mini20}). For \CSII\
the hyperfine structure was taken into account in the fitting procedure
by fixing the frequency separations and intrinsic relative LTE intensities
between the hyperfine components, and assuming the same line width for
all of them (see Ungerechts et al.~\cite{unge86} for a more detailed description
of the method).
The different $K$ components of \MAC\ where fitted simultaneously by
fixing their frequency separations to the laboratory values and assuming
the same FWHM for all of them. In Fig.~\ref{fctts} we plot the maps of
the parameters derived from the fit to the \CSII(2--1) line, namely the
opacity of the main hyperfine transition, the peak LSR velocity, and the
line width. Figure~\ref{fmac} shows the same quantities for the \MAC(6--5)
transition, with the only difference that the integral under the $K$=0
and 1 components is plotted instead of the line opacity.

Both molecules clearly trace the HMC, where the line emission and optical
depth attains their maximum, as well as the \MAC\ line width. One can also
see that the velocity gradually changes across the HMC, a result consistent
with the well known NE--SW velocity gradient observed in several studies
(see e.g. Beltr\'an et al.~\cite{belt18} and references therein) and
interpreted as rotation of the core. We note that this velocity gradient
seems to be part of the global velocity field of the region. In fact,
on a larger scale the molecular gas appears blue-shifted to the SSW
and red-shifted to the NNE, consistent with the findings of Beltr\'an
et al.~(\cite{belt22b}).

\begin{figure}
\centering
\resizebox{8.5cm}{!}{\includegraphics[angle=0]{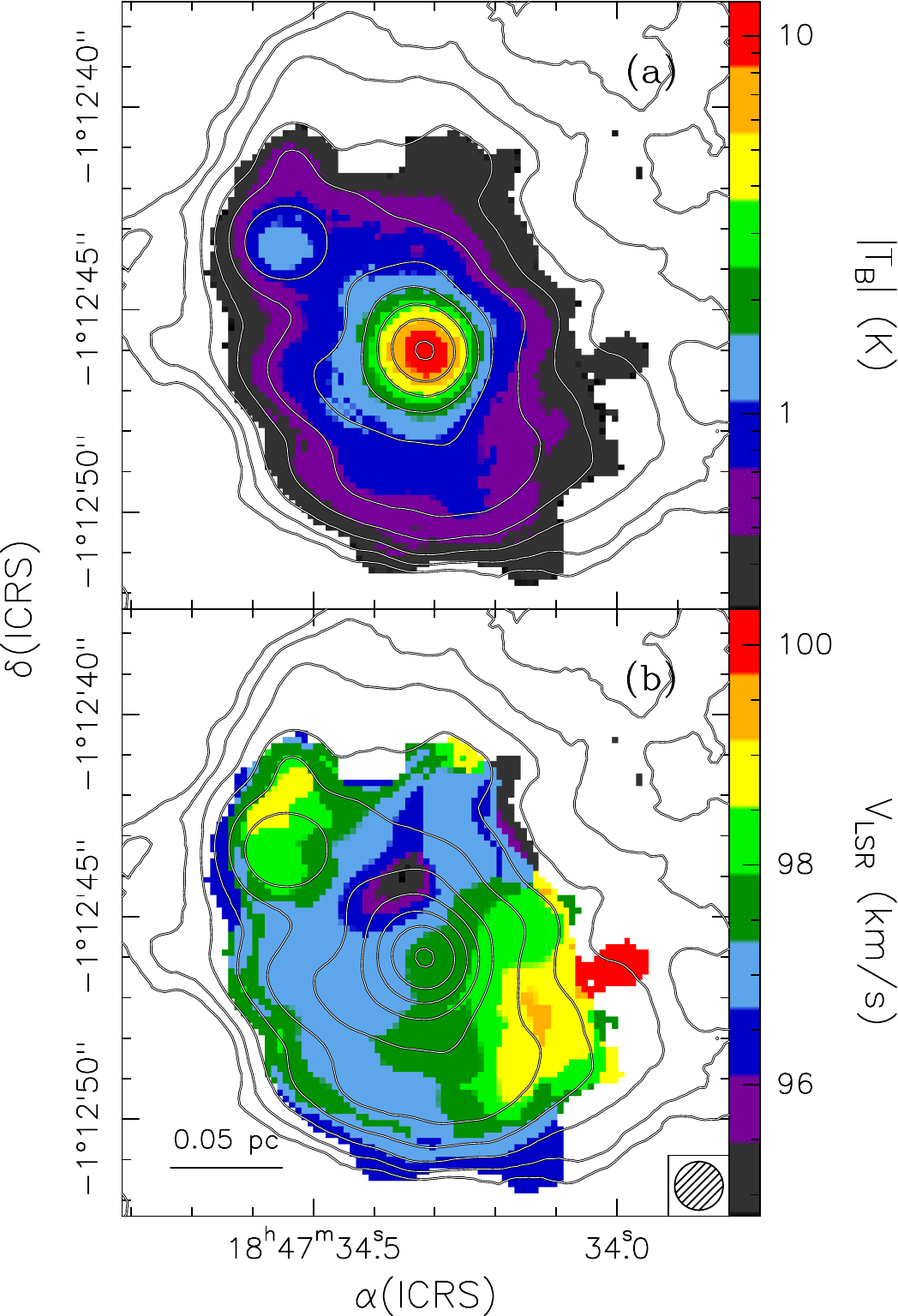}}
\caption{
{\bf a.} Map (colour image) of the absolute value of the minimum (negative)
brightness temperature of the $N$=1--0, $J$=1/2--1/2, $F$=3/2--3/2 hyperfine
component of the CN(1--0) line.
{\bf b.} Map of the LSR velocity at which the minimum brightness shown in
panel a is attained.  The synthesised beam is shown in the bottom right
corner. The contour map represents the 3~mm continuum emission and is
the same as in Fig.~\ref{fvmaps}c.
}
\label{fcna}
\end{figure}

The CN(1--0) transition was used to trace the gas seen in absorption against
the 3~mm continuum. This is useful to establish the motion of the molecular
gas with respect to the ionised gas, as the material seen in absorption is
that lying between the \HII\ region and the observer. Given the complex
hyperfine structure of the transition, for our purposes we used only
the $N$=1--0, $J$=1/2--1/2, $F$=3/2--3/2 component at 113191.2787~MHz,
as it is not blended with other lines. In Fig.~\ref{fcna}a we show the
map obtained by taking the minimum intensity at each pixel of the map,
while the corresponding LSR velocity is shown in Fig.~\ref{fcna}b. The
two maps are compared to the map of the 3~mm continuum emission and one
sees that two absorption dips are found toward the \UC\ region and the HMC,
namely towards the continuum peaks. The most red-shifted absorption is found
towards the NE and SW of the \HII\ region, with the region lying between
the HMC and the \UC\ region being the most blue-shifted. A comparison of
the velocity fields of the molecular and ionised components is presented
in Sect.~\ref{skin}.

\begin{figure}
\centering
\resizebox{8.5cm}{!}{\includegraphics[angle=0]{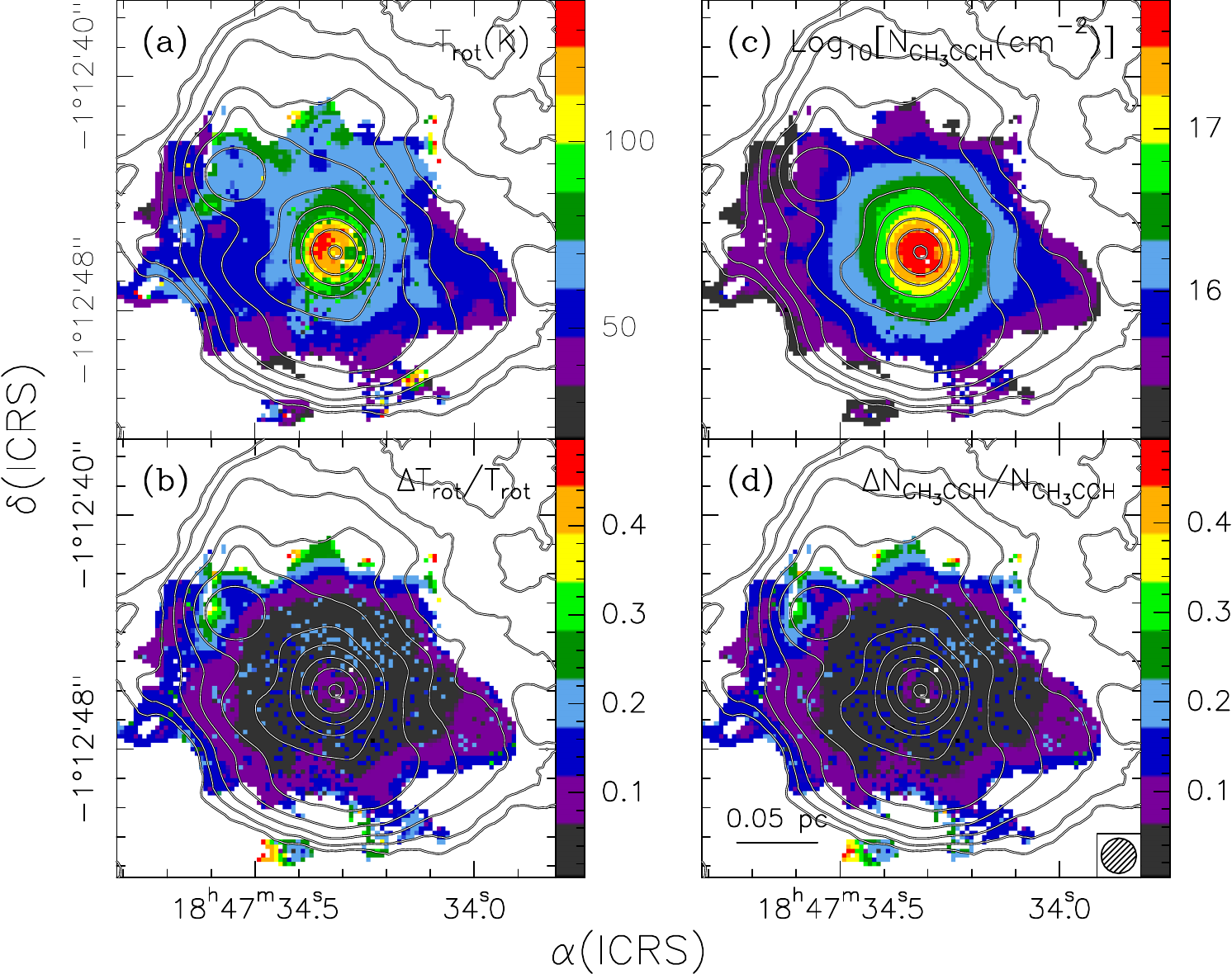}}
\caption{
{\bf a.} Map (colour image) of the rotational temperature obtained from
the rotation diagram of the \MAC\ molecule. Contours represent the same
map of the 3~mm continuum emission as in Fig.~\ref{fvmaps}c.
{\bf b.} Map of the relative error on the rotation temperature.
{\bf c.} Same as panel a, for the \MAC\ column density.
{\bf d.} Same as panel b, for the \MAC\ column density.
The synthesized beam is shown in the bottom right corner.
}
\label{fmtn}
\end{figure}

From the integrated intensity maps of the detected \MAC(6--5) $K$ components we
could obtain column density (\Ncol) and rotation temperature (\Trot) maps
by fitting rotation diagrams pixel by pixel, under the LTE approximation
(see, e.g., Fontani et al.~\cite{font02}). The method was applied only to
the spectra where at least four (out of six) $K$ lines were detected. The
results are shown in Fig.~\ref{fmtn} and compared to the map of the 3~mm
continuum emission. As expected, both quantities reach their maximum at
the position of the HMC, with a temperature of $\sim$140~K and a \MAC\
column density of $\sim6\times10^{17}$~\cmq. The mean value of \Trot\ over
the HMC is $\sim$90~K, significantly less than that of $\sim$150--165~K
estimated by Beltr\'an et al.~(\cite{belt18}) towards the center of the HMC
(see their Tables~5 and~7). Such a difference is not surprising because
these authors observed higher-excitation lines of a less abundant molecule
(\MCN) with better angular resolution ($\sim$0\farcs2), which makes their
observations sensitive to more internal and hence hotter parts of the molecular
core.

\subsubsection{Mass from continuum emission}
\label{scont}

From our ALMA data we can also obtain an estimate of the mass of the
molecular gas enshrouding the \HII\ region. For this purpose we need to
subtract the free-free continuum contribution from the 3~mm continuum
map of Fig.~\ref{fvmaps}c. One could extrapolate the 1~cm map of
Fig.~\ref{fvmaps} to 3~mm and subtract this from the ALMA map. However,
the residual image contains substantial negative structures, most likely
due to the different {\it uv} coverages of the two maps. We thus prefer to
adopt the method proposed by Cesaroni et al.~(\cite{cesa23}), who express
the ratio between the free-free and dust emissions as a function of the
ratio between two observing frequencies and the corresponding spectral
index. The assumption is that both the free-free and dust emissions are
optically thin, as expected at a wavelength of 3~mm.

\begin{figure}
\centering
\resizebox{8.5cm}{!}{\includegraphics[angle=0]{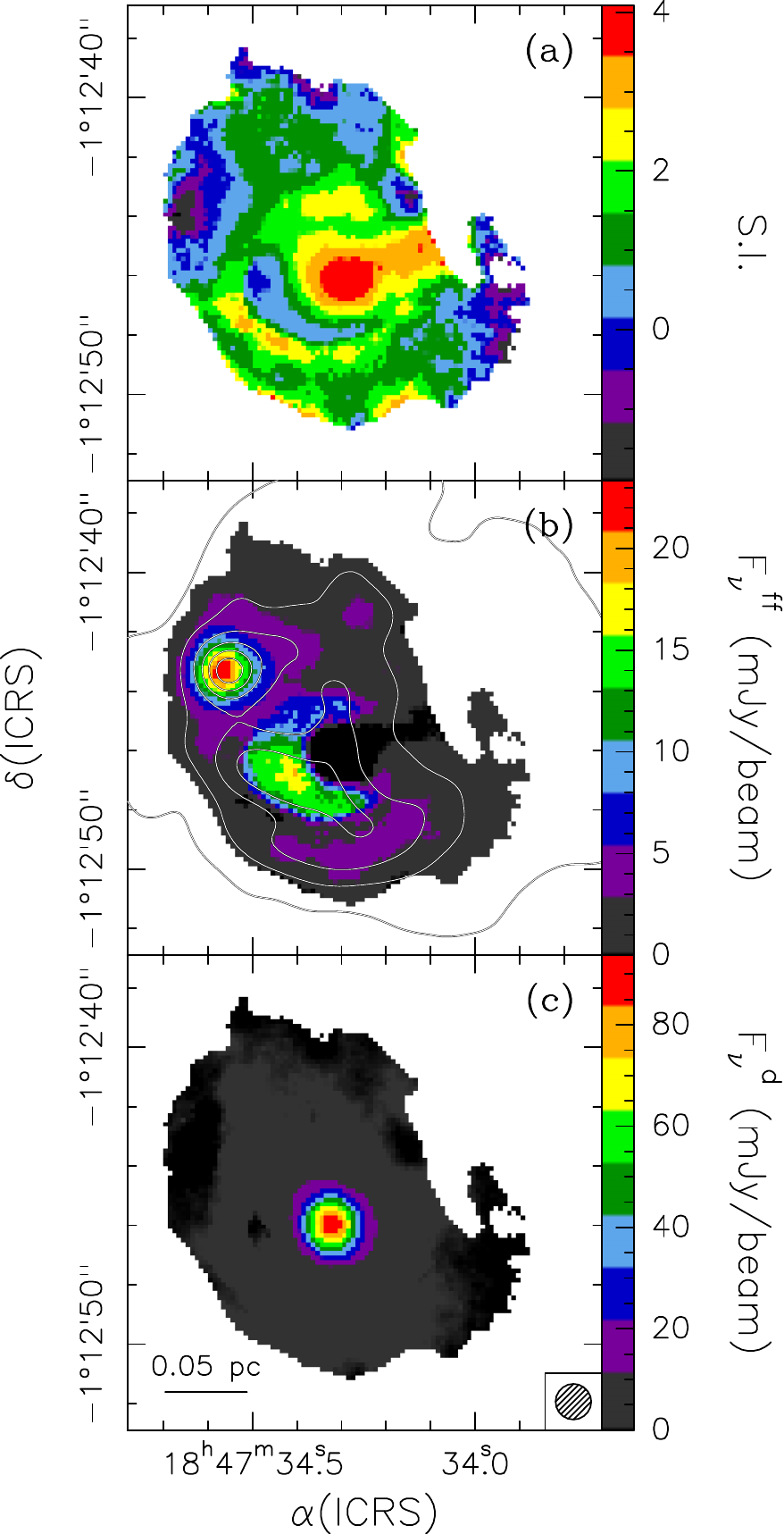}}
\caption{
{\bf a.} Map (colour image) of the spectral index of the 3~mm continuum
emission computed from the emission at the minimum and maximum frequencies
observed in the GUAPOS survey.
{\bf b.} Contour map of the 1~cm continuum emission overlaid on the map
of the 3~mm free-free continuum emission calculated using the spectral
index information (see text). Contour levels range from 1 to 26 in steps
of 5~mJy/beam.
{\bf c.} Same as panel b, for the 3~mm dust continuum emission.
The synthesised beam is shown in the bottom right corner.
}
\label{fsimap}
\end{figure}

In our case, we used the GUAPOS maps at the lowest (84.2~GHz) and highest
(115.8~GHz) frequencies. The spectral index was computed from the ratio
between the two maps and the correction was obtained by applying Eq.~(3)
of Cesaroni et al.~(\cite{cesa23}) pixel by pixel.
This method allows to separate the contribution of free-free emission
to the total flux from that of thermal dust emission, using measurements
at two frequencies.
Figure~\ref{fsimap}
shows the spectral index map and the computed maps of pure free-free and thermal
dust emission. When applying Eq.~(3) of Cesaroni et al.~(\cite{cesa23}),
we chose a fiducial dust spectral index of 3, corresponding to a dust opacity index of 1.
A few values of the map in Fig.~\ref{fsimap}a happen to fall below --0.1
(pure free-free emission) or above +3 (pure dust emission). In these cases
the spectral index was reset respectively to --0.1 and +3.

The morphology of the free-free map is qualitatively similar to that of the
1~cm map (Fig.~\ref{fvmaps}), as one can see from the comparison between
these two maps shown in Fig.~\ref{fsimap}b. This gives us confidence with
the method used. The dust map peaks at the HMC position, consistent with
the continuum images obtained at higher frequencies (see e.g. Beltr\'an et
al.~\cite{belt21}) where the free-free contribution is negligible.

\begin{figure}
\centering
\resizebox{8.5cm}{!}{\includegraphics[angle=0]{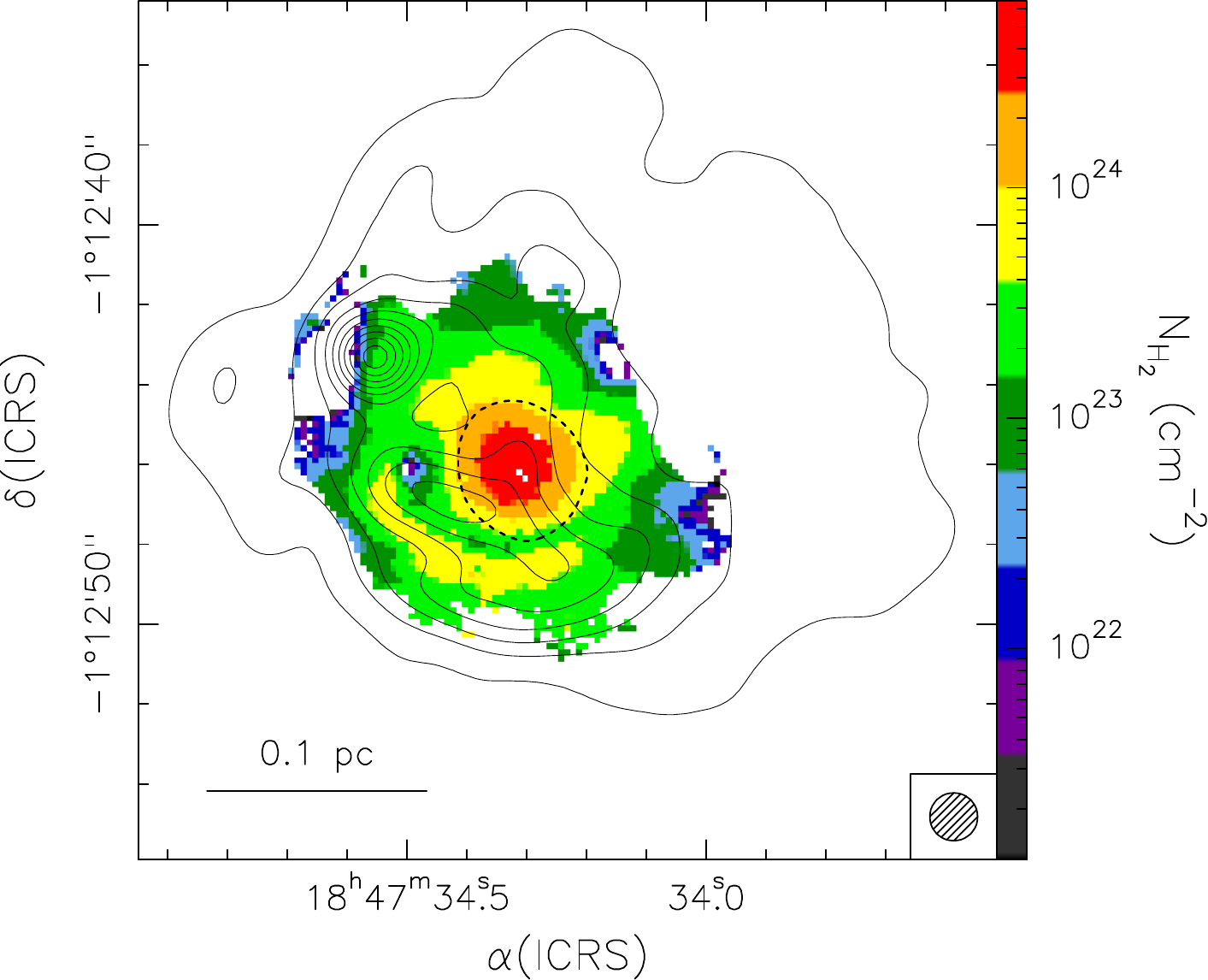}}
\caption{
Map of the \HM\ column density obtained from the dust emission map
in Fig.~\ref{fsimap}c and the temperature map in Fig.~\ref{fmtn}a. The
contours represent the map of the 1~cm continuum emission and are the
same as in Fig.~\ref{fpeaks}. The dotted pattern outlines the approximate
border of the HMC.  The synthesised beam is shown in the bottom right corner.
}
\label{fnchm}
\end{figure}

Now, we can estimate the mass of the neutral gas surrounding the \HII\
region from the dust emission map in Fig.~\ref{fsimap}c and the temperature
map obtained from \MAC\ (Fig.~\ref{fmtn}a). Knowledge of the flux and
temperature of the dust allows us to calculate a column density map (see
Fig.~\ref{fnchm}) using, e.g., Eq.~(2) of Schuller et al.~(\cite{schul09}),
where we assume temperature equilibrium between dust and gas, a gas-to-dust
mass ratio of 100, a mean molecular weight of 2.8, and a dust absorption
coefficient $\kappa(\nu)=1~{\rm cm^2~g^{-1}}~(\nu/230.6~{\rm GHz})$. To
calculate the mass of the clump around the \HII\ region we integrated over
the whole map in Fig.~\ref{fnchm} after removing the contribution of the
HMC\footnote{For this purpose we defined the border of the HMC as the 10\%
contour level of the \CSII(2--1) integrated emission, represented by
the dotted pattern in Fig.~\ref{fvmaps}}, as we are interested only in
the large scale structure. The result is $\sim$110~\Msun, which is to
be taken as a lower limit to the mass of the whole clump enshrouding the
\HII\ region because part of the 3~mm continuum emission is resolved out
by ALMA. In fact, Cesaroni et al.~(\cite{cesa91}) estimated a clump mass
of $\sim$290--1100~\Msun\ (after re-scaling their values to a distance of
3.75~kpc) from their single-dish observations of \CSI.

\begin{figure}
\centering
\resizebox{8.5cm}{!}{\includegraphics[angle=0]{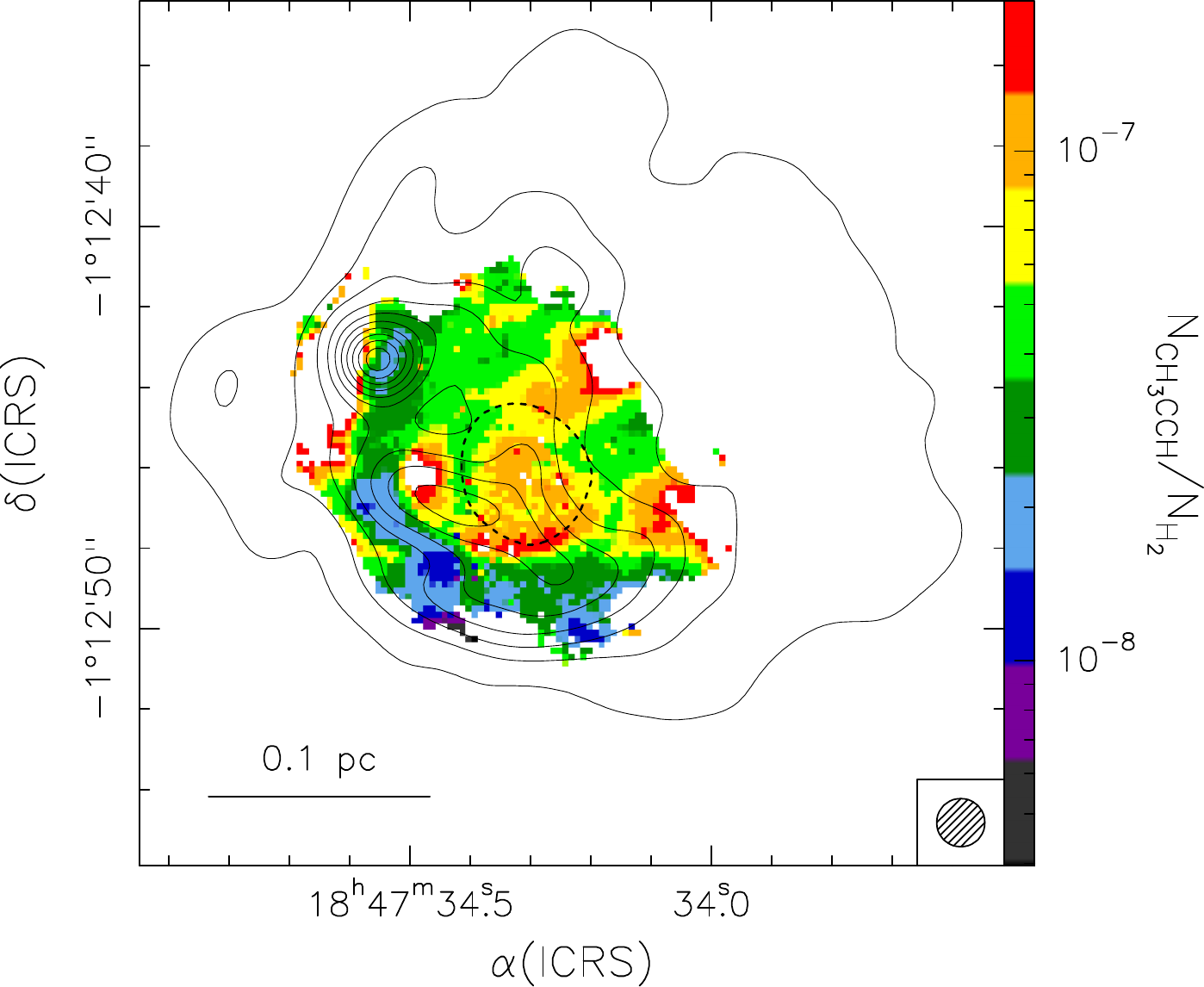}}
\caption{
Map of the \MAC\ abundance relative to \HM.
The contours represent the map of the 1~cm
continuum emission and are the same as in Fig.~\ref{fpeaks}.
The dotted pattern outlines the approximate border of the HMC.
The synthesised beam is shown in the bottom right corner.
}
\label{fxmac}
\end{figure}

Finally, from the ratio between the column density maps in Figs.~\ref{fmtn}c
and~\ref{fnchm} we could also obtain a map of the \MAC\ abundance with
respect to \HM\ ($X_{\rm CH_3CCH}$). This is shown in Fig.~\ref{fxmac}. We
can infer that $X_{\rm CH_3CCH}$ at the position of the HMC is on average
$8.2\times10^{-8}$, slightly higher than the mean value outside the HMC
($5.5\times10^{-8}$).

\section{Discussion}
\label{sdis}

In the following we use our findings to draw some conclusion on
the 3D structure of the \HII\ region and associated molecular clump.

\subsection{Pressure balance}

The morphology of the \G\ region at radio and IR wavelengths (see
Fig.~\ref{fhigal}) suggests a scenario where the ionised gas is confined to
the east by the molecular gas and is expanding approximately to the NW. It
is thus interesting to establish whether pressure balance exists at the
interface between the ionised gas and the molecular gas. Such a balance
is described by the expression
\begin{equation}
 2 \, n_{\rm e} \, T_{\rm e} = \frac{N_{\rm H_2}}{\Delta z} \, T_{\rm k}
  \label{epress}
\end{equation}
where $n_{\rm e}$ and $T_{\rm e}$ are the electron number density and
temperature, $N_{\rm H_2}$ and $T_{\rm k}$ the column density and kinetic
temperature of the molecular gas, and $\Delta z$ the length of the molecular
gas along the line of sight.
We have ignored the contribution of the magnetic field
pressure in this expression. This pressure can be estimated from the expression
$B^2/(8\pi k)$, where $B$ is the magnetic field strength and $k$ the
Boltzmann constant. According to Law et al.~(\cite{law25}), the $B$-field
intensity in this region lies in the range 0.04--0.09~mG; hence the
maximum magnetic pressure is $\sim$2$\times$$10^6$~\cmc\,K, while the
pressure inside the \HII\ region is
$2 \, n_{\rm e} \, T_{\rm e}$$\simeq$$2\times10^7$--$10^8$~\cmc\,K.
We conclude that the B-field pressure is negligible.

Our aim is to establish if pressure equilibrium can exist at the
interface between the \HII\ region and the molecular clump, namely
if Eq.~(\ref{epress}) is satisfied. This can be checked by deriving
the value of $\Delta z$ (the only unknown quantity) from this equation.
The expectation is that the length of the region along the line of
sight is of the same order as that in the plane of the sky, i.e. a few
0.1~pc.

\begin{figure}
\centering
\resizebox{8.5cm}{!}{\includegraphics[angle=0]{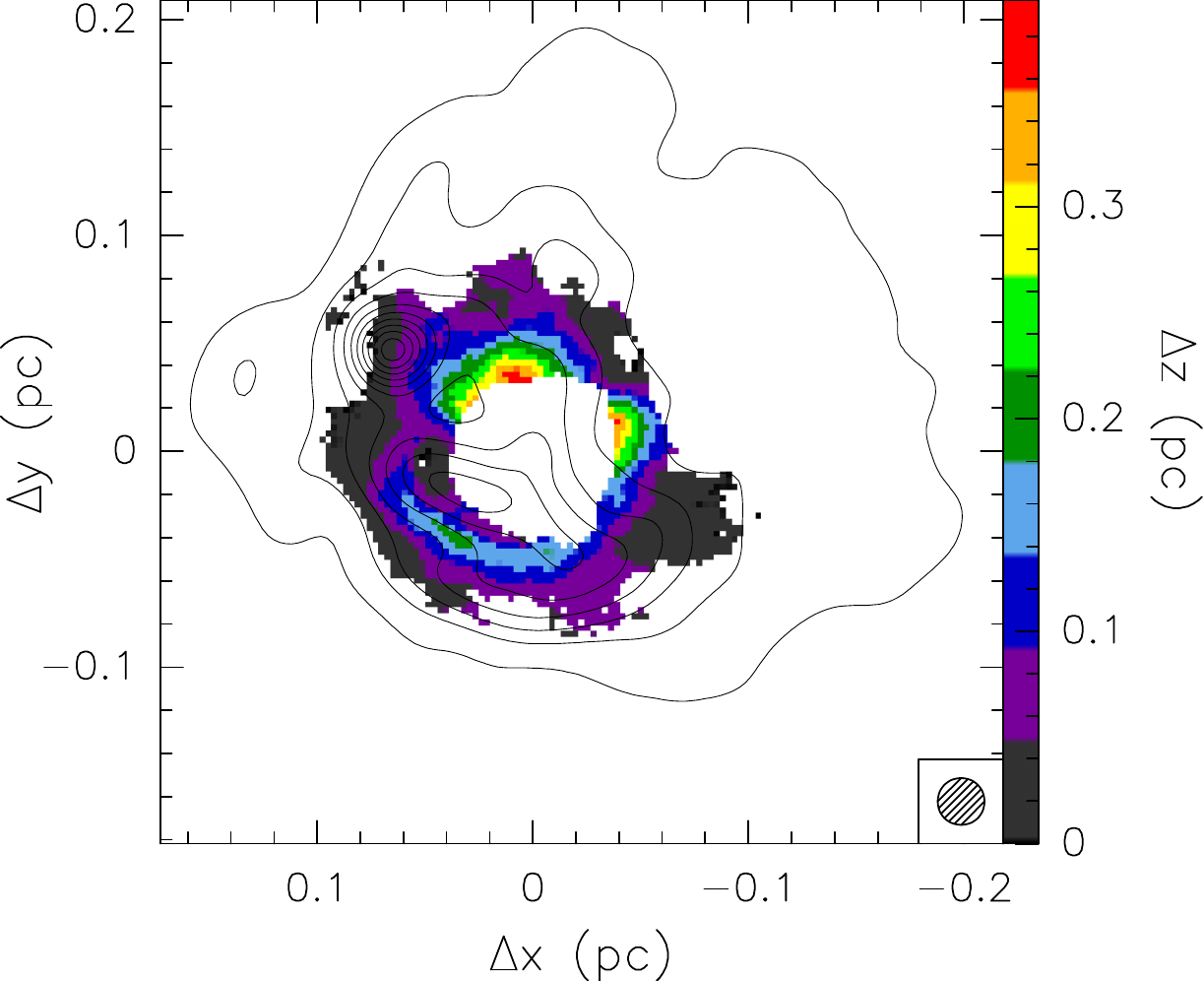}}
\caption{
Map of the estimated geometrical thickness of the molecular gas along the
line of sight. The RA and Dec axes are expressed in pc to ease the
comparison with $\Delta z$. The 0,0 position corresponds to the phase
center of the ALMA observations. The contours
represent the map of the 1~cm continuum emission and are the same as in
Fig.~\ref{fpeaks}. The synthesised beam is shown in the bottom right corner.
}
\label{fdz}
\end{figure}

From Eq.~(\ref{epress}) one can express $\Delta z$ as a function of the other
four parameters, whose values we obtain from Figs.~\ref{fnete}, \ref{fmtn}a,
and~\ref{fnchm}. The map of $\Delta z$ is shown in Fig.~\ref{fdz},
where the mean value is $\sim$0.1~pc, which confirms our expectation.
We conclude that pressure confinement is likely to be at work at the
interface between the \HII\ region and the molecular cloud.

\subsection{Kinematics of the region}
\label{skin}

\begin{figure}
\centering
\resizebox{8.5cm}{!}{\includegraphics[angle=0]{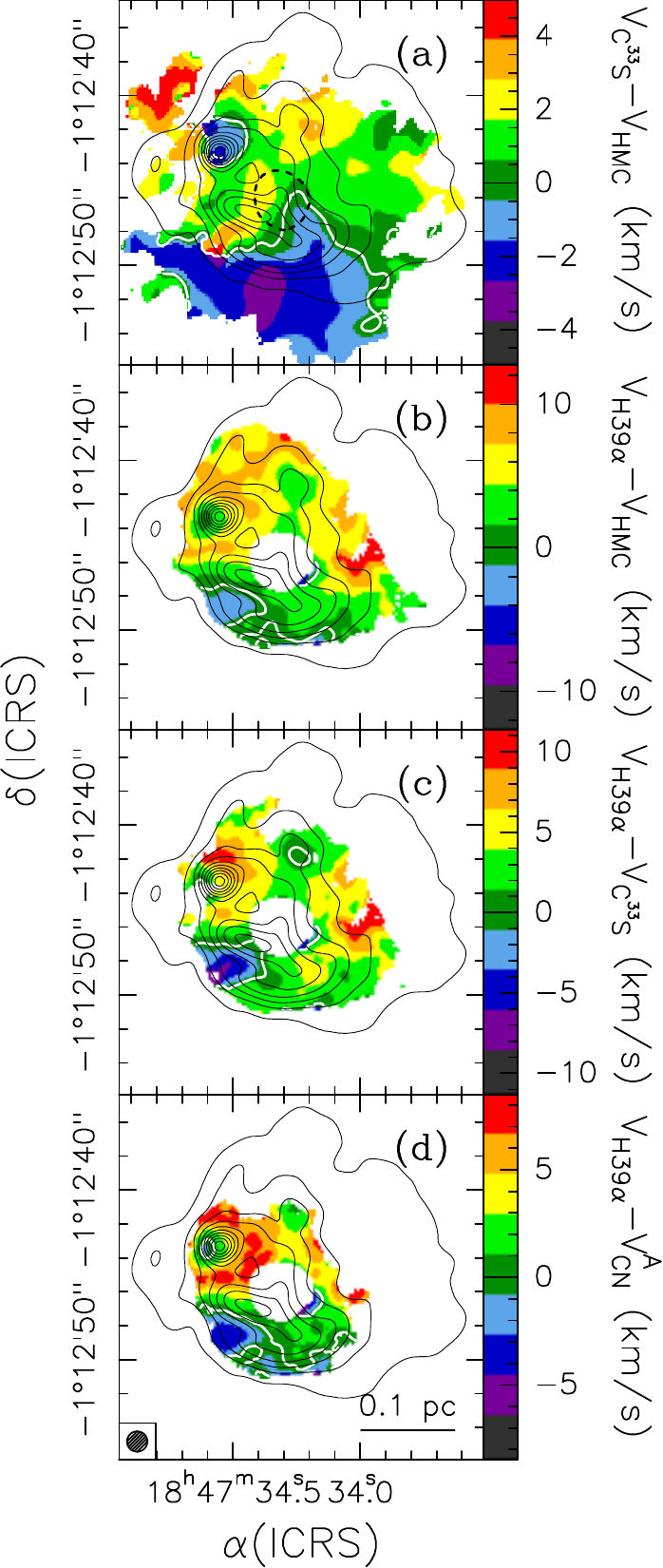}}
\caption{
{\bf a.} Map of the difference between the velocity of the \CSII(2--1)
line and the systemic LSR velocity of the HMC (96.5~\kms). The white
contours correspond to 0~\kms.  The black contours are the same as in
Fig.~\ref{fpeaks}. The dotted pattern outlines the approximate border of
the HMC.
{\bf b.} Same as panel a, for the difference between the velocity of the
H39$\alpha$ line and that of the HMC.
{\bf c.} Same as panel a, for the difference between the velocity of the
H39$\alpha$ line and that of the \CSII(2--1) line.
{\bf d.} Same as panel a, for the difference between the velocity of the
H39$\alpha$ line and that of the absorption dip of the CN(1--0) line.
the HMC.
The synthesised beam is shown in the bottom right corner.
}
\label{fdv1}
\end{figure}

It is interesting to compare the velocity field of the molecular gas
with that of the ionised gas. First of all, we consider the velocities of
the \CSII(2--1) and H39$\alpha$ lines with respect to the systemic LSR velocity
of the HMC (96.5~\kms). The corresponding maps are shown in Figs.~\ref{fdv1}a
and~\ref{fdv1}b. For a more detailed comparison, in Fig.~\ref{fdv1}c we plot
the difference between the recombination line velocity and the \CSII(2--1)
line velocity at the same position. While the ionised gas is red-shifted
all over the \HII\ region, with the sole exception of a small area to the SE,
the molecular gas is blue-shifted to the south and towards the \UC\ region
and is close to the systemic velocity over the remaining area.
Our hypothesis is that the ionised gas is expanding away from
the observer, which in turn suggests that the molecular gas associated
with the HMC is located mostly between the \HII\ region and the observer.
As for the blue-shifted \CSII\ emission over the southern part of the map
in Fig.~\ref{fdv1}c, it is likely due to overlap with another molecular
cloud that Beltr\'an et al.~(\cite{belt22b}) propose to be colliding with
the cloud containing the \HII\ region and the HMC.

To establish the 3D distribution of the different gas components and thus
confirm our hypothesis that the ionised gas is expanding away from the
observer and from the molecular clump enshrouding the HMC,
we take advantage of the presence of absorption in the CN(1--0) line.
In Fig.~\ref{fdv1}d we show the difference between the velocity of the
H39$\alpha$ line and that of the absorption dip in the CN line. The map is
qualitatively very similar to that in Fig.~\ref{fdv1}c, which proves that
the CN absorption and \CSII\ emission trace the same region. Therefore,
since the CN line is seen in absorption, the corresponding gas must be
located between the observer and the \HII\ region, with the latter
expanding away from the molecular clump in a ``champagne flow'' (see Yorke
et al.~\cite{yor83} and references therein).

\begin{figure}
\centering
\resizebox{8.5cm}{!}{\includegraphics[angle=0]{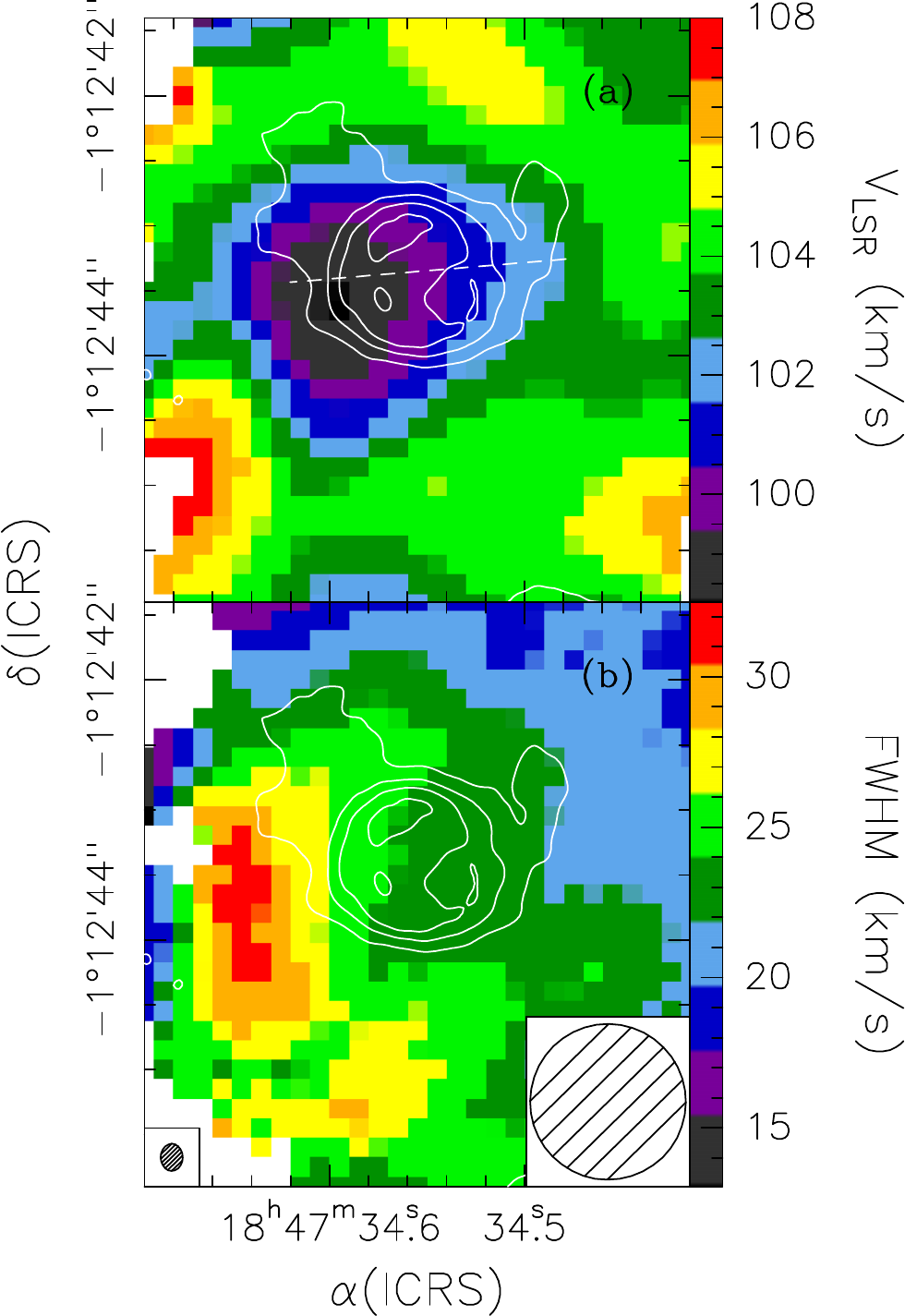}}
\caption{
{\bf a.} Contour map of the 7~mm continuum emission (enlargement of the map in
Fig.~\ref{fvmaps}) overlaid on the H40$\alpha$ velocity map (colour image)
of Fig.~\ref{frvlsr}. The dashed line marks the approximate axis of the
cometary shaped \UC\ region as well as the approximate direction of the velocity
gradient across it.
{\bf b.} Same as panel a, for the line FWHM.  The synthesised beams are
shown in the bottom left (for the 7~mm map) and right (for the velocity
map) corners.
}
\label{fvmm}
\end{figure}

Finally, it is worth analysing the velocity field towards the \UC\ region.
The inverse P-Cygni profiles of Fig.~\ref{fspeca} hints at the
existence of residual infall of the surrounding molecular gas. The presence
of strong interaction between the molecular and the ionised gas through
the eastern border of the \UC\ region is witnessed by the kinematics
of the ionised component. In Fig.~\ref{fvmm} we compare maps of the LSR
velocity and FWHM of the H40$\alpha$ line to a map of the 7~mm continuum
emission. The peak of the \UC\ region is skewed to the east, namely towards
the molecular gas, so that the ionised gas has a cometary shape hinting
at expansion towards the west. In particular, from Fig.~\ref{fvmm}a one
sees that across the \UC\ region the ionised gas has a velocity gradient
almost in the same direction as the symmetry axis of the cometary shape
(see the dashed line in the figure)
and the velocity over the whole \UC\ region is red-shifted with respect
to that of the HMC (96.5~\kms). Further evidence of interaction between
the ionised gas with the molecular cloud is provided by Fig.~\ref{fvmm}b
where the maximum FWHM of the recombination line is found right at the
eastern border of the \UC\ region. Noticeably, this is also the location
where the electron temperature peaks in Fig.~\ref{fnete}a, which lends
further support to the existence of strong activity in this area.

\subsection{A 3D view of the \G\ region}
\label{sview}

Based on the analysis of the previous sections, we propose a tentative
3D picture of the \HII\ region and the molecular clump enshrouding it.
In summary, from the previous discussion we conclude that:
\begin{itemize}
\item The HMC and the lower-density molecular gas have basically the
 same velocity (Fig.~\ref{fdv1}a), which suggests that the HMC is lying
 inside the molecular clump.
\item The CN absorption feature towards the \HII\ region proves that
 the medium-density gas lies between the ionised gas and the observer.
\item The ionised gas (including the \UC\ region) is red-shifted with
 respect to the molecular gas (Figs.~\ref{fdv1}b and~\ref{fdv1}c), which
 proves that the \HII\ region is expanding away from the observer to the NW.
\item The red-shifted absorption towards the \UC\ region (see the inverse
 P-Cygni profiles in Fig.~\ref{fspeca}) proves the existence of residual
 infall.
\item The NE--SW velocity gradient across the HMC (Figs.~\ref{fctts}b
 and~\ref{fmac}b) is consistent with that observed inside the HMC, which
 suggests that the lower-density gas enshrouding the HMC could be undergoing
 rotation too.
\end{itemize}

\begin{figure}
\centering
\resizebox{8.5cm}{!}{\includegraphics[angle=0]{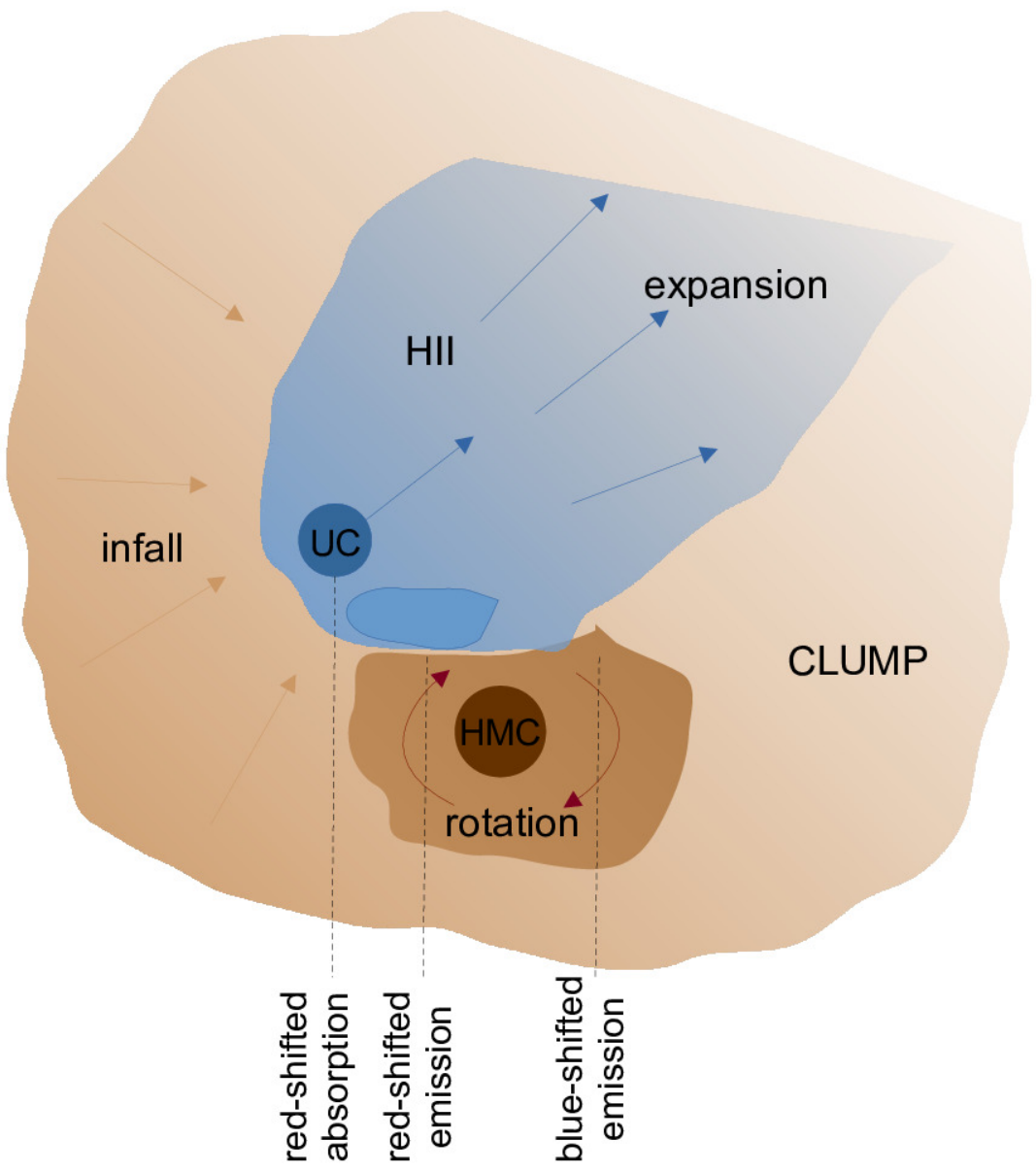}}
\caption{
Sketch of the proposed scenario for the \G\ star forming region. The dashed
lines represent the three lines of sight corresponding to the spectra in
Figs.~\ref{fspeca}--\ref{fspecc}. The observer is located at the bottom
of the figure.
}
\label{fview}
\end{figure}

With all the above in mind, we propose a scenario that should explain
the observed features. This is depicted in Fig.~\ref{fview}. The basic
idea is that there are two star formation centers in \G, one associated
with the \HII\ region and another, less evolved, associated with the HMC.
Both are located inside a molecular clump whose remnant is still clearly
visible on the parsec scale at 8~\mic\ (see Fig.~\ref{fhigal}a), as a
dark lane extending to the east and to the south. This medium-density
gas is still infalling as witnessed by the red-shifted absorption seen
along the line of sight through the \UC\ region. The presence of infall
is consistent with the scenario proposed by Beltr\'an et al.~(\cite{belt22b}),
where the material is accreting through filaments onto a ``hub'' where the
\HII\ region and HMC are located. The high-density gas more
tightly associated with the HMC is rotating, as well as the HMC itself, while
the ionised gas is expanding to the NW in a champagne flow receding from
the observer. We believe that the HMC lies in front of the \HII\ region,
because the medium-density gas is rotating with the HMC and is seen in
(slightly red-shifted) absorption (Fig.~\ref{fspecb}). Moreover, the HMC
should not be too far from the \HII\ region surface, because we see an
electron density enhancement between the HMC and the SE border of the
\HII\ region (see Fig.~\ref{fnete}c); this hints at a proximity between
the two objects.

\section{Summary and conclusions}

The present study is part of the GUAPOS project, which observed the \G\
HMC and associated \UC\ region using the whole bandwidth of the 3~mm ALMA
receivers with 1\farcs2 angular resolution. We used the data of 15 hydrogen
recombination lines and several molecular transitions to investigate the
velocity field and physical structure of the ionised and molecular gas.
The recombination line maps were fitted with a suitable non-LTE model to
derive the electron temperature and density all over the \HII\ region,
while the same parameters for the molecular gas were obtained from rotation
diagrams of the \MAC(6--5) line emission. We find that the \HII\ region
is expanding to the NW, as suggested also by its cometary shape, while
it is pressure confined to the SE. Comparison between the velocity fields
obtained from the recombination and molecular lines suggests that the
molecular gas is still undergoing residual infall towards the peak of
the \UC\ region and is expanding away from it in the southern part.
Based on our results, we propose a 3D picture where most of the molecular
gas is distributed to the east of the \HII\ region and between the latter and
the observer, with the HMC lying close to the interface between the ionised
gas and the lower-density molecular gas. While in all likelihood the
\HII\ region is ionization bound, we believe that, due to the cometary shape,
a non negligible fraction of the stellar photons longward of 912~\AA\ might
escape from the surrounding molecular cloud, thus leading to an underestimation
of the bolometric luminosity of the region.

\begin{acknowledgements}
R.C., M.T.B., and A.L. acknowledge financial support through the INAF
Large Grant ``The role of MAGnetic fields in MAssive star formation'' (MAGMA).
C.M. acknowledges funding from the European Research Council (ERC) under
the European Union's Horizon 2020 program through the ECOGAL Synergy
grant (ID~855130) and funding from INAF Mini Grants RSN2 2024 ``Zodyac''
CUP C83C25000340005.
V.M.R. and L.C. acknowledge support from the grant PID2022-136814NB-I00 by the
Spanish Ministry of Science, Innovation and Universities/State Agency
of Research MICIU/AEI/10.13039/501100011033 and by ERDF, UE; V.M.R. also
acknowledges the grant
RYC2020-029387-I funded by MICIU/AEI/10.13039/501100011033 and by ``ESF,
Investing in your future'', and from the Consejo Superior de Investigaciones
Cient{\'i}ficas (CSIC) and the Centro de Astrobiolog{\'i}a (CAB) through
the project 20225AT015 (Proyectos intramurales especiales del CSIC); and
from the grant CNS2023-144464 funded by MICIU/AEI/10.13039/501100011033
and by ``European Union NextGenerationEU/PRTR''.
A.S.-M. acknowledges support from the RyC2021-032892-I grant funded
by MCIN/AEI/10.13039/501100011033 and by the European Union `Next
GenerationEU’/PRTR, as well as the program Unidad de Excelencia María
de Maeztu CEX2020-001058-M, and support from the PID2023-146675NB-I00
(MCI-AEI-FEDER, UE).
A.L.-G. acknowledges support from the grant PID2022-136814NB-I00 by the
Spanish Ministry of Science, Innovation and Universities/State Agency of
Research MICIU/AEI/10.13039/501100011033 and by ERDF, UE; and from the
Consejo Superior de Investigaciones Cient{\'i}ficas (CSIC) and the Centro
de Astrobiolog{\'i}a (CAB) through the project 20225AT015 (Proyectos
intramurales especiales del CSIC).
The project that gave rise to these results received the support of a fellowship from the ``la Caixa'' Foundation (ID 100010434). The
fellowship code is LCF/BQ/PR25/12110012.
This paper makes use of the following ALMA data: ADS/JAO.ALMA\#2017.1.00501.S.
ALMA is a partnership of ESO (representing its member states), NSF (USA)
and NINS (Japan), together with NRC (Canada), NSC and ASIAA (Taiwan), and
KASI (Republic of Korea), in cooperation with the Republic of Chile. The
Joint ALMA Observatory is operated by ESO, AUI/NRAO and NAOJ.
This study is also based on observations made under projects 16A-181 and
23A-066 with the VLA of NRAO. The National Radio Astronomy Observatory is
a facility of the National Science Foundation operated under cooperative
agreement by Associated Universities, Inc.
\end{acknowledgements}

\begin{appendix}

\section{Radius of \HII\ region}
\label{arad}

\begin{figure}
\centering
\resizebox{8.5cm}{!}{\includegraphics[angle=0]{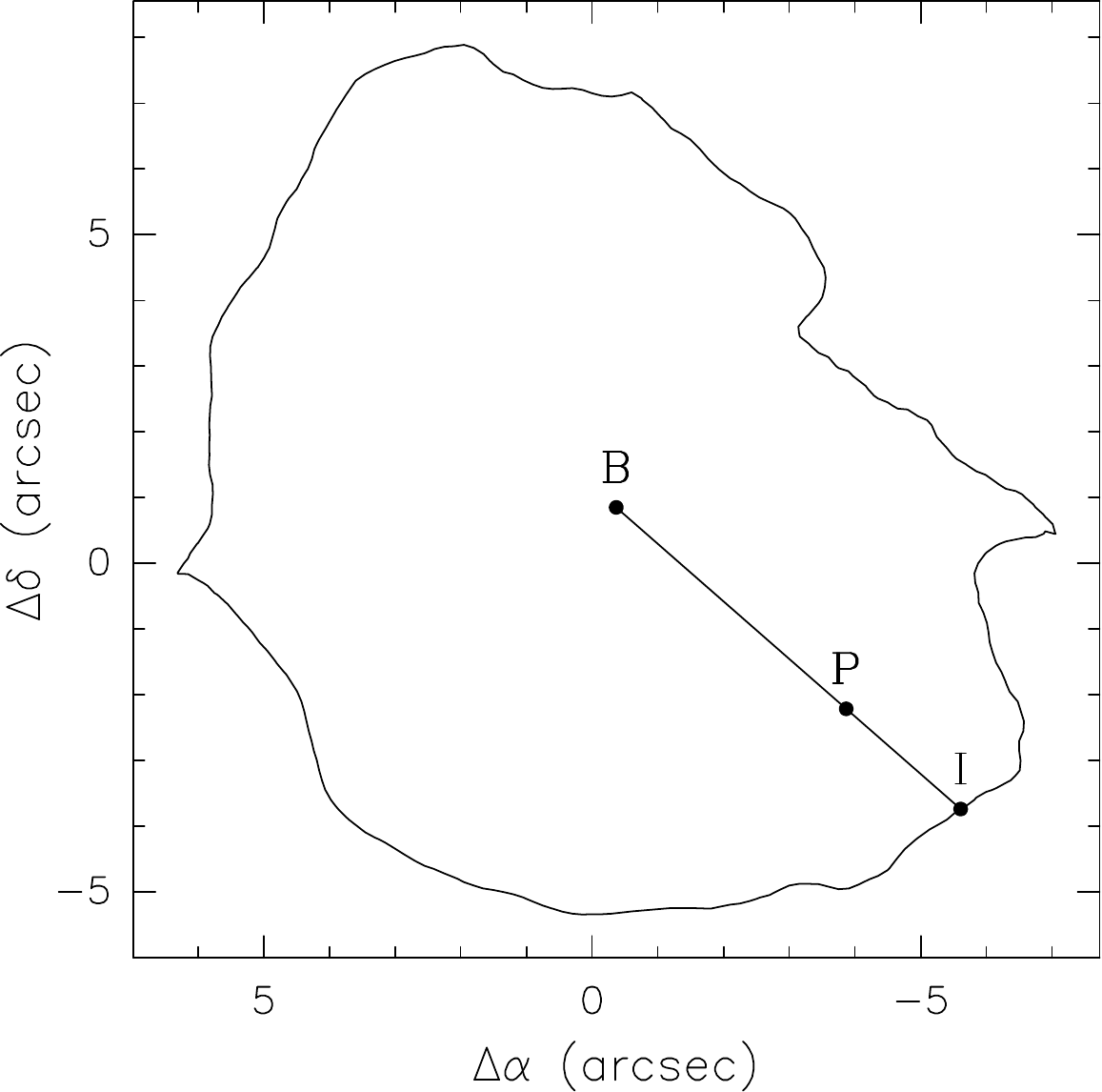}}
\caption{
Sketch of the method adopted to estimate the \HII\ region radius for a generic
point P. The contour corresponds to the 5$\sigma$ level of the 3~mm continuum
emission and is assumed to be the border of the \HII\ region, while B is the
barycenter of this contour. I is the intersection between the half-line originating
in B and passing through P. The segment BI is the radius used in our model
calculations to determine the intensity of the hydrogen recombination lines in P.
The 0,0 position is the phase center of the ALMA observations.
}
\label{fsk}
\end{figure}

As explained in Sect.~\ref{smod}, our model assumes that the \HII\ region
is spherical and requires knowledge of its radius. Since the \G\ \HII\
region is far from an ideal sphere, we have decided to estimate a value of
the radius that depends on the position. In Fig.~\ref{fsk} we show a sketch
that illustrates our approach. We assume that the border of the \HII\ region
coincides with the 5$\sigma$ contour level of the 3~mm continuum emission
and the \HII\ region center is taken as the barycenter, B, of such a contour, whose
coordinates are
\begin{eqnarray}
x_{\rm B} & = & \frac{\sum_{i=1}^N\,x_i}{N} \\
y_{\rm B} & = & \frac{\sum_{i=1}^N\,y_i}{N}
\end{eqnarray}
with $i$ indicating a generic point of the border and $N$ is the number of
these points.  For a generic point, P, of the \HII\ region we assume that
the \HII\ region radius to be used as an input for our model is equal to the
segment BI that connects the center, B, with the intersection, I, between the
\HII\ region border and the half-line originating in B and passing through P.
Obviously, this definition implies that the radius varies with P but
we believe that this mirrors the irregular shape of the \HII\ region more
reliably than approximating the \HII\ region with a sphere of a given radius.
The values of the radius obtained in this way range from 3\farcs8 to 7\farcs6.

\end{appendix}

\end{document}